\pdfoutput=1

\documentclass[11pt,twoside,a4paper,cmspaper,final,collab]{cms-tdr}
\def\svnVersion{2465ef0-D}\def\svnDate{2019/08/09}\def\cmsCernNoTag{CERN-EP-2019-093}\def\cmsCernDate{\today}\def\cmsMessage{"Published in the Journal of High Energy Physics as \href{http://dx.doi.org/10.1007/JHEP07(2019)150}{\doi{10.1007/JHEP07(2019)150}.}"}

\begin{document}\cmsNoteHeader{SUS-17-007}

\hyphenation{had-ron-i-za-tion}
\hyphenation{cal-or-i-me-ter}
\hyphenation{de-vices}
\RCS$HeadURL$
\RCS$Id$
\newlength\cmsTabSkip\setlength{\cmsTabSkip}{1ex}
\providecommand{\PSGcmpDo}{\ensuremath{\widetilde{\chi}^{\mp}_{1}}\xspace}
\providecommand{\CL}{CL\xspace}
\newcommand{\pttrigmiss}{\ensuremath{p_{\text{T,trig}}^{\text{miss}}}\xspace}
\newcommand{\httrigmiss}{\ensuremath{H_{\text{T,trig}}^{\text{miss}}}\xspace}
\newcommand{\mjj}{\ensuremath{m_\text{jj}}\xspace}

\cmsNoteHeader{SUS-17-007}

\title{Search for supersymmetry with a compressed mass spectrum in the vector boson fusion topology with 1-lepton and 0-lepton final states in proton-proton collisions at $\sqrt{s}=13\TeV$}

\date{\today}

\abstract{
A search for supersymmetric particles produced in the vector boson fusion topology in proton-proton collisions is presented. The search targets final states with one or zero leptons, large missing transverse momentum, and two jets with a large separation in rapidity. The data sample corresponds to an integrated luminosity of 35.9\fbinv of proton-proton collisions at $\sqrt{s}=13\TeV$ collected in 2016 with the CMS detector at the LHC. The observed dijet invariant mass and lepton-neutrino transverse mass spectra are found to be consistent with the standard model predictions. Upper limits are set on the cross sections for chargino (\PSGcpmDo) and neutralino (\PSGczDt) production with two associated jets. For a compressed mass spectrum scenario in which the \PSGcpmDo and \PSGczDt decays proceed via a light slepton and the mass difference between the lightest neutralino \PSGczDo and the mass-degenerate particles \PSGcpmDo and \PSGczDt is 1\,(30)\GeV, the most stringent lower limit to date of 112\,(215)\GeV is set on the mass of these latter two particles. }

\hypersetup{
pdfauthor={CMS Collaboration},
pdftitle={Search for SUSY in VBF with 2016 data},
pdfsubject={CMS},
pdfkeywords={CMS, supersymmetry}}

\maketitle

\section{Introduction}
\label{sec:intro}

Supersymmetry (SUSY)~\cite{PhysRevD.3.2415, JETP.13.1971, FERRARA1974413, WESS197439, PhysRevLett.49.970, BARBIERI1982343, PhysRevD.27.2359} is a theory that can simultaneously describe the particle 
nature of dark matter (DM) and solve the gauge hierarchy problem of the standard model (SM).  However, for all of its attractive features, there is as yet no direct evidence to support this theory. 
The masses of the strongly produced gluinos (\PSg) as well as the squarks (\PSQ) of the first and second generations have been excluded below approximately 2\TeV in certain simplified model 
scenarios \cite{Sirunyan:2017kqq,Sirunyan:2017cwe, ATLAScolored1, ATLAScolored2, Alwall:2008ag, Alves:2011wf}. On the other hand, the values of the masses of the weakly produced charginos 
($\PSGcpm_{\mathrm{i}}$) and neutralinos  ($\PSGcz_{\mathrm{i}}$)   are less constrained at the CERN LHC where these particles have much smaller production cross sections. The chargino-neutralino 
sector of SUSY plays an important role in establishing a connection between SUSY models and DM. The lightest neutralino \PSGczDo, as the lightest supersymmetric particle (LSP), is the canonical DM 
candidate in $R$-parity conserving SUSY extensions of the SM~\cite{Farrar:1978xj}.

\begin{figure}[tbh!]
\centering
  \includegraphics[width=0.45\textwidth]{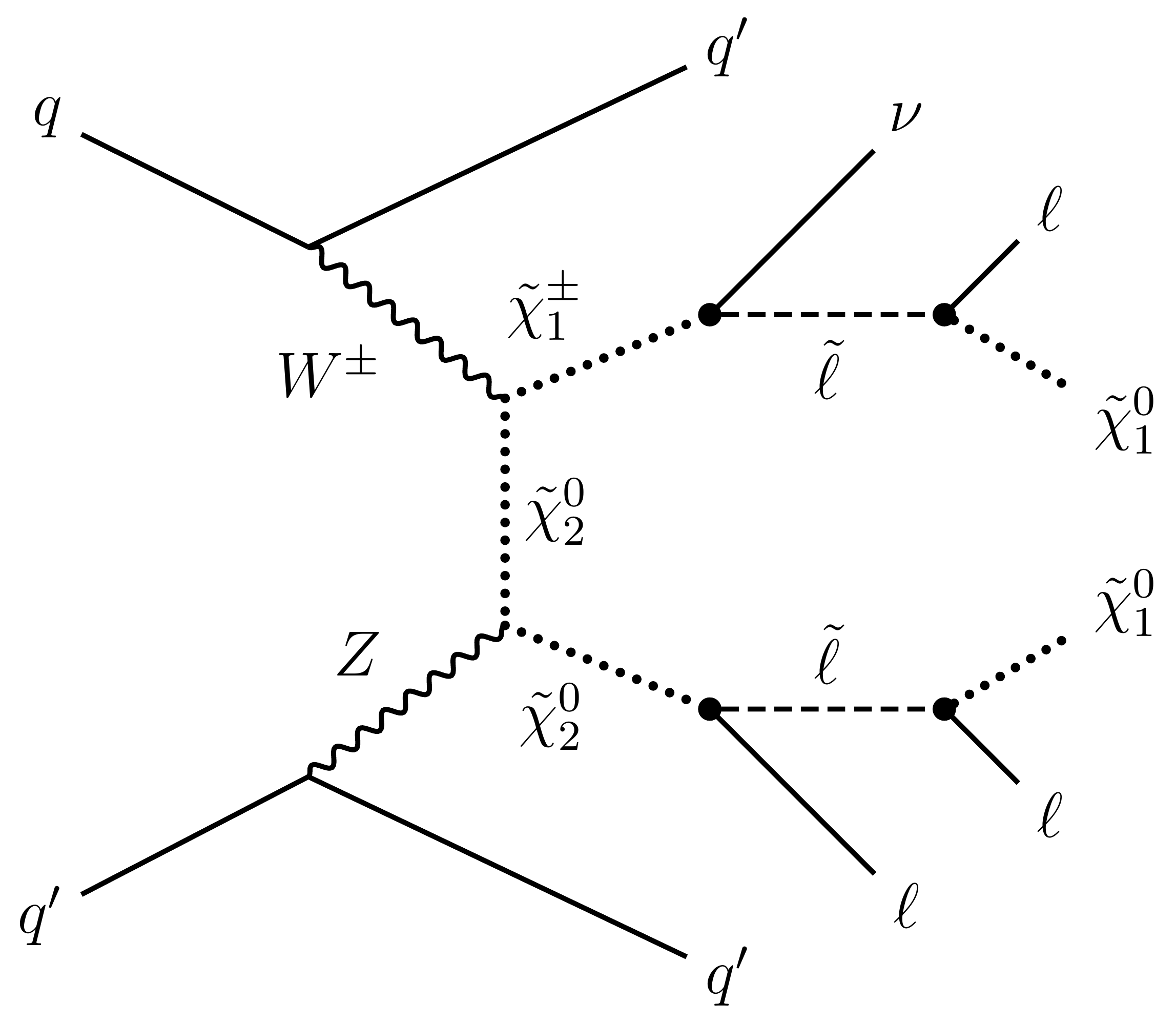}
  \includegraphics[width=0.45\textwidth]{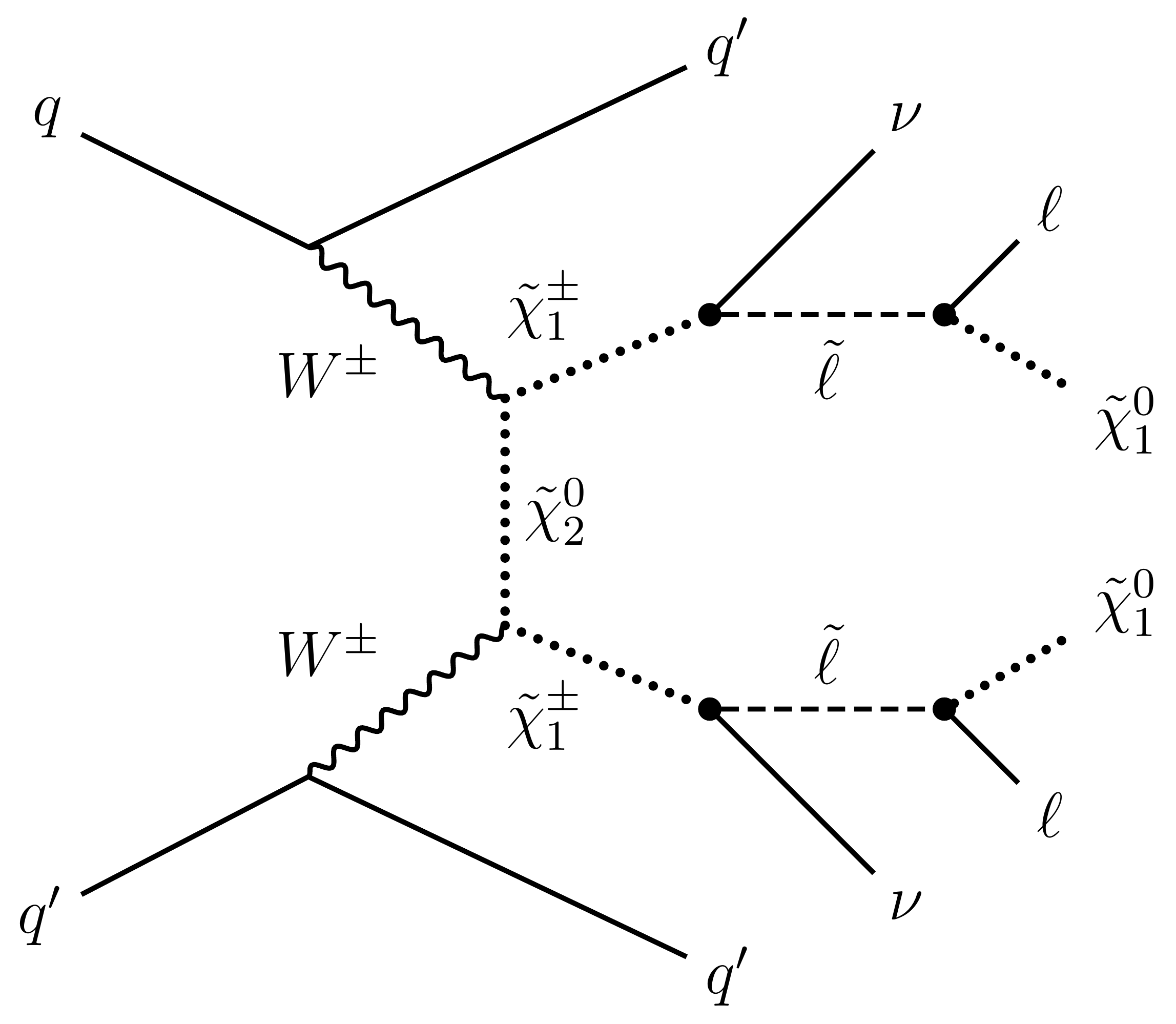}\\
  \includegraphics[width=0.45\textwidth]{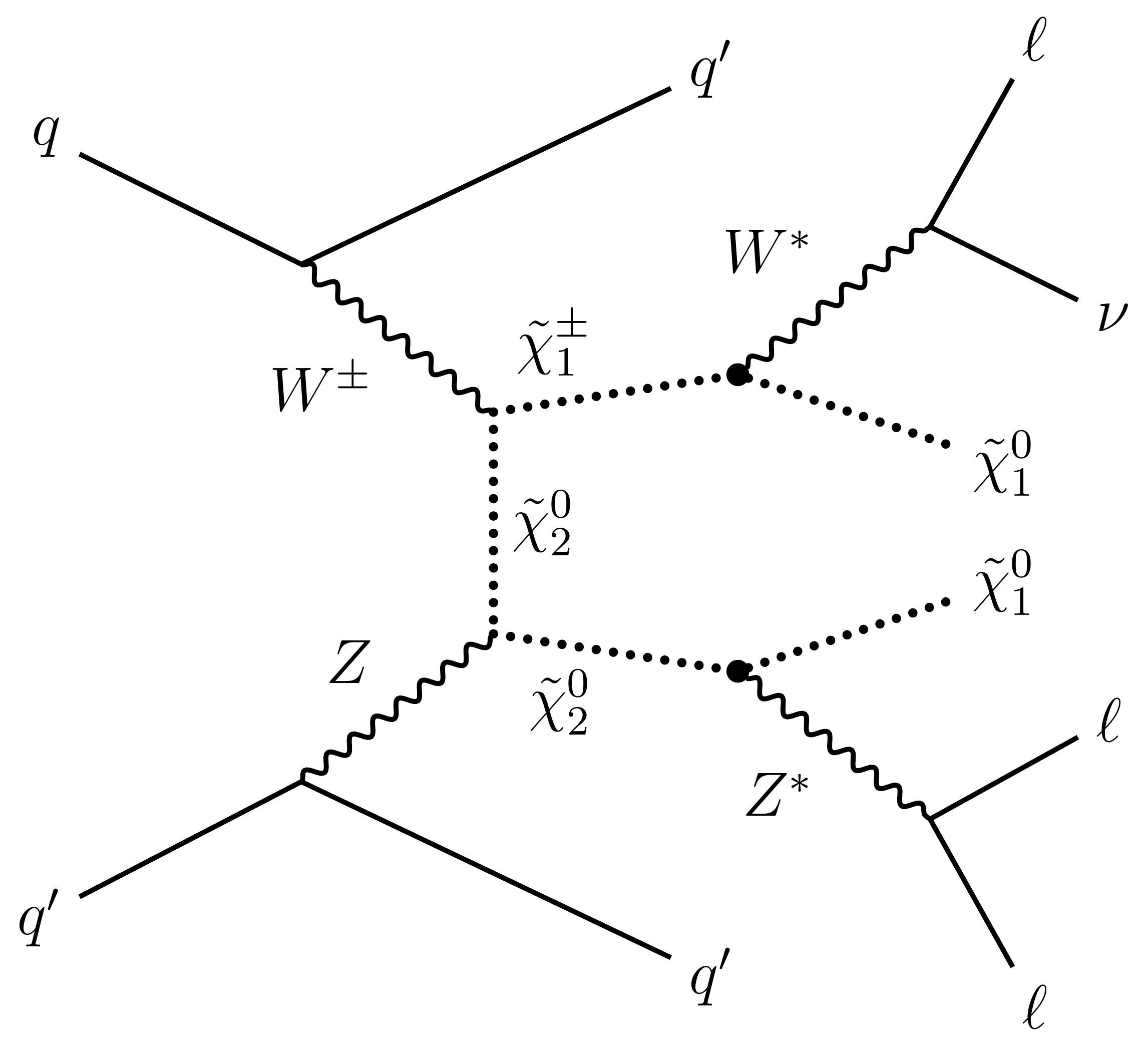}
  \includegraphics[width=0.45\textwidth]{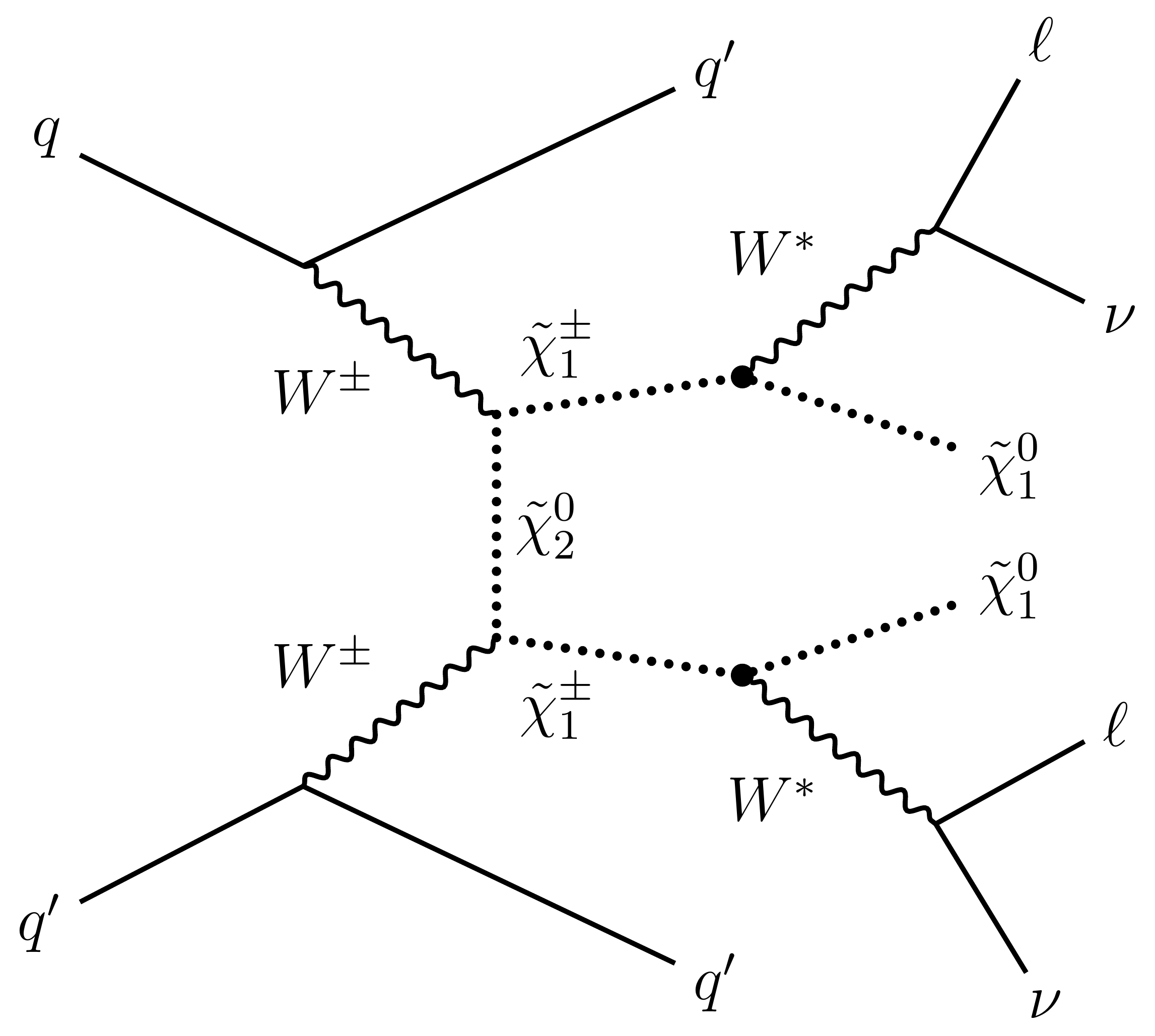}
\caption{Representative Feynman diagrams of (left) chargino-neutralino and (right) chargino-chargino pair production through vector boson fusion, followed by their decays to
leptons and the LSP \PSGczDo via a light slepton (top row) or a $\PW^{*}/\PZ^{*}$ (bottom row). Although these representative diagrams show multiple leptons in the final state, the compressed
mass spectra scenarios of interest result in low-\pt leptons, making it unlikely to reconstruct and identify more than one lepton.}
\label{fig:feynVBF}
\end{figure}

A common strategy to search for charginos and neutralinos is through Drell--Yan (DY) production processes of order $\alpha_{\mathrm{EW}}^{2}$ (electroweak coupling squared) 
involving virtual {\PW}  and {\PZ} bosons ($\PW^{*}/\PZ^{*}$), $\cPq\cPaq'\to \PW^{*} \to \PSGcpm_{\mathrm{i}} \PSGcz_{\mathrm{j}}$, followed by their decay to final states with one or more charged leptons 
($\ell$) and missing transverse momentum (\ptmiss). These processes can include, for example, \PSGcpmDo\PSGczDt pair production followed by $\PSGcpmDo\to\ell^{\pm} \nu_{\ell} \PSGczDo$ and  
$\PSGczDt \to \ell^{\pm}  \ell^{\mp}  \PSGczDo$ via virtual SM bosons or a light slepton $\widetilde{\ell}$, where \PSGcpmDo (\PSGczDt) is the lightest (next-to-lightest) chargino (neutralino), and 
where the LSP \PSGczDo is presumed to escape without detection leading to significant missing momentum. However, these searches are experimentally difficult in cases where the mass of the LSP is only slightly less than the masses of other charginos and neutralinos, making these so-called compressed spectrum scenarios important search targets using new techniques.
While the exclusion limits in Refs.~\cite{CMSEWK,Aad:2014nua,ATLASewk} can be as stringent as $m_{\PSGcpmDo} < 650\GeV$ for a massless \PSGczDo, they weaken to only approximately 100\GeV for $\Delta 
m \equiv m_{\PSGcpmDo} - m_{\PSGczDo} = 2\GeV$, assuming decays of the \PSGcpmDo and \PSGczDt to leptonic final states proceed through the mediation of virtual {\PW} and {\PZ} bosons~\cite{ATLASTsoftdilepton,CMSsoftdilepton}. As the mass difference between SUSY particles decreases, the momenta available to the co-produced SM particles are small, resulting in ``soft'' decay products having low transverse momentum (\pt). Therefore, the traditional searches using DY processes suffer in the compressed spectrum scenarios since the SM particles used for discrimination become more difficult to reconstruct as their momenta decrease. In contrast, chargino and neutralino production via vector boson fusion (VBF) processes of order $\alpha_{\mathrm{EW}}^{4}$ are very useful in tackling these interesting compressed SUSY scenarios~\cite{VBFPheno}.
In VBF processes, electroweak SUSY particles are pair-produced in association with two high-\pt oppositely-directed jets close to the beam axis (forward), resulting in a large dijet invariant mass 
(\mjj). The use of two high-\pt VBF jets in the event topology effectively suppresses the SM background while, simultaneously, creating a recoil effect that facilitates both the detection of \ptmiss 
in the event and the identification of the soft decay products in compressed-spectrum scenarios because of their natural kinematic boost~\cite{PRD812010115011,VBFPheno2}. 
Figure~\ref{fig:feynVBF} shows the Feynman diagrams for two of the possible VBF production processes: chargino-neutralino and chargino-chargino production.

The CMS collaboration reported the first results of a SUSY search using the VBF dijet topology for charginos and neutralinos in the minimal supersymmetric standard model (MSSM), using a data sample corresponding to an integrated luminosity of 19.7\fbinv  of proton-proton collision data at $\sqrt{s}=8\TeV$~\cite{Khachatryan:2015kxa}.
That analysis considered SUSY models with light staus (\PSgt) leading to leptonic decay modes of the charginos and neutralinos (\eg, $\PSGczDt\to\tau^{\pm}\PSgt^{\mp}\to\TT\PSGczDo$). In the presence of a light slepton, it is likely that \PSGcpmDo decays to $\ell^{\pm}\nu_{\ell} \PSGczDo $ and \PSGczDt decays to $\ell^{+}\ell^{-} \PSGczDo$.
Thus, charginos and neutralinos were probed using final states with two leptons and two additional jets consistent with the VBF topology. In the compressed mass spectrum scenario, where the mass difference between the \PSGczDo and $\PSGczDt/\PSGcpmDo$ particles was taken to be 50\GeV, \PSGczDt and \PSGcpmDo masses below 170\GeV were excluded.

In this paper, a search is presented for the electroweak production of SUSY particles in the VBF topology using data collected in 2016 with the CMS detector and corresponding to an integrated luminosity of 35.9\fbinv of proton-proton collisions at a center-of-mass energy of $\sqrt{s}=13\TeV$. Besides the two oppositely directed forward jets ($\mathrm{j}$) that define the VBF configuration, the search requires the presence of zero or one soft lepton and large \ptmiss. The events are classified into two categories based on the lepton content, $0\ell \mathrm{jj}$ and $1\ell \mathrm{jj}$, with the latter having three different final states:
$\Pe \mathrm{jj}$, $\mu \mathrm{jj}$, and $\tauh \mathrm{jj}$, where \tauh denotes a hadronically decaying {\Pgt} lepton. The $0\ell \mathrm{jj}$ final state (also referred to as the ``invisible'' channel) provides the best sensitivity to the $\Delta m < 10\GeV$ scenarios, where the leptons from the $\PSGczDt/\PSGcpmDo$ decays are ``lost'', either because their momenta are too low to reconstruct or because they fail to satisfy the identification requirements. The soft single-lepton channels were not utilized in the 8\TeV search and thus this analysis extends the previous search performed only in the two-lepton final state.
The dijet invariant mass distribution \mjj is the sensitive variable used to discriminate possible SUSY signal from background in the $0\ell \mathrm{jj}$ channel, while the transverse mass \mT between the lepton and \ptmiss is used in the $1\ell \mathrm{jj}$ channels.

The backgrounds are evaluated using data wherever possible. The general strategy is to define control regions, each dominated by a different background process and with negligible contamination from signal events, through modification of the nominal selection requirements. These control regions are used to measure the \mjj and \mT shapes and probabilities for background events to satisfy the VBF selection requirements. If the background contribution from a particular process is expected to be small or if the above approach is not feasible, the \mjj and \mT shapes are taken from simulation. In these cases, scale factors, defined as the ratio of efficiencies measured in data and simulation, are used to normalize the predicted rates to the data.

The paper is organized as follows. The CMS detector is described in Section \ref{sec:detector}. The reconstruction of electrons, muons, \tauh leptons, jets, and \ptmiss is presented in Section \ref{sec:obj}.
The simulated SUSY signal and background samples are discussed in Section \ref{sec:samples}, followed by the description of the event selection in Section \ref{sec:sel} and the background estimation in Section \ref{sec:bkg}.
Systematic uncertainties are summarized in Section \ref{sec:sys}, and the results are presented in Section \ref{sec:res}. Section \ref{sec:sum} contains a summary of the paper.

\section{The CMS detector}
\label{sec:detector}

The central feature of the CMS apparatus is a superconducting solenoid of 6\unit{m} internal diameter, providing a magnetic field of 3.8\unit{T}. Located within the solenoid volume are silicon pixel and strip detectors, a lead tungstate electromagnetic calorimeter (ECAL), and a brass and scintillator hadron calorimeter (HCAL). Muons are measured in gas-ionization detectors embedded in the steel flux-return yoke outside the solenoid.
Extensive forward calorimetry complements the barrel and endcap detectors by covering the pseudorapidity range $3.0 < \abs{\eta} <5.2$.

The inner silicon tracker measures charged tracks with $\abs{\eta} < 2.5$ and provides an impact parameter resolution of approximately 15\mum and a transverse momentum resolution of about 1.5\% for 100\GeV charged particles. Collision events of interest are selected using a two-tiered trigger system.
The first level trigger (L1), composed of custom hardware processors, selects events at a rate of around 100\unit{kHz}.
The second level trigger, based on an array of microprocessors running a version of the full event reconstruction software optimized for fast processing, reduces the event rate to around 1\unit{kHz} before data storage.
A detailed description of the CMS detector, along with a definition of the coordinate system and relevant kinematic variables, can be found in Ref.~\cite{Chatrchyan:2008zzk}.

\section{Event reconstruction and particle identification}
\label{sec:obj}

The particle-flow (PF) algorithm is used to reconstruct the jets and \ptmiss used in this analysis~\cite{pflow}.
The PF technique combines information from different subdetectors to produce a mutually-exclusive collection of particles (namely muons, electrons, photons,
charged hadrons, and neutral hadrons) that are used as input for the jet clustering algorithms.
The missing transverse momentum vector \ptvecmiss is defined as the negative
vector sum of the momenta of all reconstructed PF candidates in an event, projected on the plane perpendicular to the beam axis.
The magnitude of \ptvecmiss is \ptmiss~\cite{CMS-PAS-JME-16-004}.
The production of undetected particles such as SM neutrinos and the SUSY \PSGczDo is inferred by the measured \ptmiss.
The reconstructed vertex with the largest value of summed physics-object $\pt^2$ is taken to be the primary $\Pp\Pp$ interaction vertex. The physics objects are the jets, clustered using the jet
finding algorithm~\cite{Cacciari:2008gp,fastjetpaper} with the tracks assigned to the vertex as inputs, and the associated missing transverse momentum, taken as the negative vector sum of the \pt
of those jets.

Jets are clustered using the \FASTJET anti-\kt algorithm~\cite{Cacciari:2008gp,fastjetpaper}, with a distance parameter of 0.4.
Only jets that satisfy the identification criteria designed to reject particles from multiple proton-proton interactions (pileup) and anomalous behavior in
the calorimeters are considered in this analysis~\cite{Cacciari:2007fd}.
The jet energy scale and resolution are calibrated through correction factors that depend on the \pt and $\eta$ of the jet \cite{CMS:JetResol}.
Jets with $\pt > 60\GeV$ have a reconstruction-plus-identification efficiency of approximately 99\%, while 90--95\% of pileup jets are rejected~\cite{CMS-PAS-JME-13-005}.
Jets originating from the hadronization of bottom quarks (\cPqb\ quark jets) are identified using the
combined secondary vertex (CSV) algorithm~\cite{heavy-flavour_jets}, which exploits observables related to the long lifetime of \PB\ hadrons.
For jets with $\pt > 20\GeV$  and $\abs{\eta} < 2.4$, the \cPqb\ tagging algorithm is operated at a working point such that the probability of correctly
identifying a \cPqb\ quark jet is approximately 60\%, while the probability of misidentifying a jet generated from a light-flavor quark or gluon as a \cPqb\ quark jet is
approximately 1\%~\cite{heavy-flavour_jets}.

Muons are reconstructed using the inner silicon tracker and muon detectors~\cite{MUONreco}.
Quality requirements based on the minimum number of measurements in the silicon
tracker, pixel detector, and muon detectors are applied to suppress
backgrounds from decays-in-flight and hadron shower remnants that reach the muon system.
Electrons are reconstructed by combining tracks produced by the
Gaussian-sum filter algorithm with ECAL clusters~\cite{Khachatryan:2015hwa}.
Requirements on the track quality, the shape of the energy deposits in the ECAL,
and the compatibility of the measurements from the tracker and the ECAL
are imposed to distinguish prompt electrons from charged pions
and from electrons produced by photon conversions.
The electron and muon reconstruction efficiencies are $>$99\% for $\pt > 8\GeV$.

The electron and muon candidates are required to satisfy isolation criteria in order to reject
non-prompt leptons from the hadronization of quarks and gluons.
Relative isolation is defined as the scalar sum of the \pt values of reconstructed charged and neutral particles within a cone of radius
$\DR \equiv \sqrt{\smash[b]{(\Delta\eta)^{2} + (\Delta\phi)^{2}}}=0.4$ (where $\phi$ is the azimuthal angle in radians) around
the lepton-candidate track, divided by the \pt of the lepton candidate.
To suppress the effects of pileup, tracks from charged particles not associated with the primary vertex are excluded from the isolation sum, and the contribution to pileup from reconstructed neutral
hadrons is subtracted~\cite{Cacciari:2007fd}.
The contribution from the electron or muon candidate is removed from the sum. The value of the isolation variable is required to be
$\le$0.0821 for electrons and $\le$0.25 for muons~\cite{MUONreco,Khachatryan:2015hwa}.

The total efficiency for the muon identification and isolation requirements is 96\%
for muons with $\pt > 10\GeV$ and $\abs{\eta} < 2.1$.
The rate at which pions undergoing $\Pgppm\to\PGmpm \Pgngm $ decay are misidentified as prompt muons is $10^{-3}$ for pions
with $\pt > 10\GeV$ and $\abs{\eta} < 2.1$.
The total efficiency for the electron identification and isolation requirements is 85 (80)\% for electrons with $\pt > 10\GeV$ in the barrel (endcap) region~\cite{Khachatryan:2015hwa}.
The jet$\to\Pe$ misidentification rate is $5\times 10^{-3}$ for jets with $\pt > 10\GeV$ and $\abs{\eta} < 2.1$~\cite{Khachatryan:2015hwa}.

Hadronic decays of {\Pgt} leptons are reconstructed and identified using the hadrons-plus-strips
algorithm \cite{1748-0221-11-01-P01019}, which is designed to optimize the performance of the \tauh reconstruction by considering specific \tauh
decay modes.
To suppress backgrounds in which light-quark or gluon jets can mimic \tauh decays, a \tauh candidate is
required to be spatially isolated from other energy deposits in the event.
The isolation variable is calculated using a multivariate boosted decision tree technique
within a cone of radius $\DR = 0.5$ around the direction of the \tauh candidate and considering the energy deposits of particles not included in the reconstruction of the \tauh decay mode.
Additionally, \tauh candidates are required to be distinguishable from electrons and muons in the
event by using dedicated criteria based on the consistency among the measurements in the tracker,
calorimeters, and muon detectors. With these requirements, the contribution from electrons and muons being misidentified as genuine \tauh candidates is negligible ($\ll$0.1\%).

The identification and isolation efficiency at the tight working point used in this analysis is approximately 50\% for a \tauh lepton with $\pt > 20\GeV$ and $\abs{\eta} < 2.1$, while
the probability for a jet to be misidentified as a \tauh is 1--5\%, depending on the \pt and $\eta$ values of the \tauh candidate~\cite{1748-0221-11-01-P01019}. Although the
tight working point is used to define the $\tauh \mathrm{jj}$ signal region, a loose working point is used to obtain multijet enriched control samples for estimation of the background rate
in the signal region. The identification and isolation efficiency for a \tauh lepton at the loose working point used in this analysis is approximately 60\%, while
the probability for a jet to be misidentified as a \tauh is about 10\%.

The event selection criteria used in each search channel are summarized in Section~\ref{sec:sel}.

\section{Signal and background samples}
\label{sec:samples}

The SM background composition depends on the final state of each channel considered in the analysis.
The main backgrounds in the four channels considered in the analysis are estimated using data-driven methods. Negligible or minor backgrounds are obtained directly from simulation.
For the $\Pe \mathrm{jj}$ and $\mu \mathrm{jj}$ channels, the main backgrounds are from \ttbar production and {\PW} boson production in association with jets ({\PW}+jets). Subdominant background sources come from single top quark,
diboson ($\PW\PW$, $\PW\PZ$, and $\PZ\PZ$, collectively referred to as VV) and {\PZ}+jets production. For the $\tauh \mathrm{jj}$ channel, the main source of background
consists of SM events only containing jets produced
via the strong interaction, referred to as quantum chromodynamics (QCD) multijet events, followed by {\PW}+jets and \ttbar production. In the $0\ell \mathrm{jj}$ channel, the main backgrounds are
$\PW/\PZ$+jets and QCD multijet events, with a minor contribution from \ttbar and diboson production.

The {\PW}+jets, \ttbar, and single top quark processes produce events with genuine leptons, \ptmiss, and jets.
The {\PZ}+jets process contributes to the background composition when one of the leptons is lost as a result of detector acceptance or inefficiencies in the reconstruction and identification
algorithms. Although jets in QCD events have a 1--5\% probability of being misidentified as a $\tauh$, the large QCD multijet production cross section results in a substantial contribution of this
background to the $\tauh \mathrm{jj}$ channel.

In the $0\ell \mathrm{jj}$ channel, the {\PZ}+jets background produces genuine \ptmiss when the {\PZ} boson decays into neutrinos.
The {\PW}+jets process also has real \ptmiss when the {\PW} boson decays leptonically, but it results in a similar $0\ell \mathrm{jj}$ final state when the lepton is not observed as a consequence of the detector acceptance or is not properly reconstructed or identified because of inefficiencies in the corresponding algorithms.
The QCD multijet events can also have significant \ptmiss from mismeasurement of jet energies.

Simulated samples of signal and background events are generated using Monte Carlo (MC) event generators. The signal event samples are generated with the {\MGvATNLO} v2.3.3
generator~\cite{Alwall:2014hca} at leading order (LO) precision, considering pure electroweak pair production of $\PSGcpmDo$ and $\PSGczDt$ gauginos ($\PSGcpmDo\PSGcpmDo$,
$\PSGcpmDo\PSGcmpDo$, $\PSGcpmDo\PSGczDt$, and $\PSGczDt\PSGczDt$) with two associated partons.
Models with a bino-like $\PSGczDo$ and wino-like $\PSGczDt$ and $\PSGcpmDo$ are considered.
The signal events are generated requiring a pseudorapidity gap $\abs{\Delta \eta} > 3.5$ between the two partons, with $\pt > 30\GeV$ for each parton.
This parton level $\abs{\Delta \eta}$ requirement is verified to provide no bias with respect to the final requirement on the reconstructed dijet pseudorapidity gap. 
The LO cross sections in this paper are obtained with these parton-level requirements. 
Note that VBF $\PSGcpmDo\PSGczDt$ production is the dominant process in the models considered, composing about 60\% of the total signal cross section, while the VBF $\PSGcpmDo\PSGcmpDo$ process is the
second-largest contribution, composing about 30\% of the total signal cross section.
The VBF $\PSGcpmDo\PSGcpmDo$ and $\PSGczDt\PSGczDt$ processes compose about 10\% of the total signal cross section.
The $\PZ / \cPgg^{*}(\to \ell^{+}\ell^{-})$+jets, $\PZ(\to \Pgn_{\ell}\Pagn_{\ell})$+jets, and $\PW(\to \ell \Pgn_{\ell})$+jets backgrounds are also simulated at LO precision
using {\MGvATNLO}, where up to four partons in the final state are included in the matrix element calculation.
The background processes involving the production of a single vector boson in association with two jets exclusively through pure electroweak interactions
are simulated at LO via {\MGvATNLO}. 
The interference effect between pure electroweak and mixed electroweak-QCD production of V+jets events has been studied and found to be small~\cite{EWKZ2j}. The effect is neglected in our 
analysis and the sum of these two samples is henceforth referred to as $\PZ$+jets. 
The QCD multijet background is also simulated at LO using {\MGvATNLO}.
Single top quark and \ttbar processes are generated at next-to-leading order (NLO) using the \POWHEG v2.0 generator~\cite{Frederix:2012dh,Nason:2004rx,Frixione:2007vw,Alioli:2010xd,Re:2010bp}.
The leading order \PYTHIA v8.212 generator is used to model the diboson processes.
The \POWHEG and \MADGRAPH generators are interfaced with the \PYTHIA v8.212~\cite{Sjostrand:2014zea} program, which is used to describe the parton shower
and the hadronization and fragmentation processes with the CUETP8M1 tune~\cite{Khachatryan:2015pea}. The NNPDF3.0 LO and NLO~\cite{Ball:2014uwa} parton distribution functions
(PDFs) are used in the event generation. Double counting of the partons generated with \MGvATNLO and \POWHEG~interfaced with \PYTHIA is removed using the MLM~\cite{Alwall:2007fs} matching scheme.
The LO cross sections are used to normalize simulated signal events, while NLO cross sections are used for simulated
backgrounds~\cite{Alwall:2014hca,Alioli:2009jeNew,Re:2010bp,Czakon:2011xx}.

For both signal and background simulated events,
additional pileup interactions are generated with {\PYTHIA} and superimposed on the primary collision process. The simulated events are reweighted
to match the pileup distribution observed in data. The background samples are processed with a detailed simulation of the CMS apparatus using the
\GEANTfour package~\cite{Geant}, while the CMS fast simulation package~\cite{FastSimulation} is used to simulate the CMS detector for the signal samples.

\section{Event selection}
\label{sec:sel}

Events are selected using a trigger with a threshold of 120\GeV on both \pttrigmiss and \httrigmiss. The
variable \pttrigmiss
corresponds to the magnitude of the vector \ptvec sum of all the PF candidates reconstructed at the trigger level, while \httrigmiss is computed as
the magnitude of the vector \ptvec sum of all jets with $\pt > 20\GeV$ and $\abs{\eta} < 5.0$ reconstructed at the trigger level. The energy fraction attributed to neutral hadrons
in these jets is required to be smaller than 0.9. This requirement suppresses anomalous events with jets originating from detector noise. To be able to use the same trigger for selecting
events in the muon control samples used for background prediction, muon candidates are not included in the \pttrigmiss nor \httrigmiss computation.
The \ptmiss threshold defining the search regions is chosen to achieve a trigger efficiency greater than 95\%.

While the compressed mass spectrum SUSY models considered in this analysis result in final states with multiple leptons~\cite{VBFPheno,VBFPheno2}, the compressed mass spectra scenarios of interest
also result in low-\pt visible decay
products, making it difficult to reconstruct and identify multiple leptons. For this reason, events are required to have zero or exactly one well-identified soft lepton.
In the $\mu \mathrm{jj}$ channel, an additional lepton veto is applied by rejecting events containing a second muon ($\pt > 8\GeV$), an electron ($\pt > 10\GeV$), or a $\tauh$
candidate ($\pt > 20\GeV$). Similarly, $\Pe \mathrm{jj}$ and $\tauh \mathrm{jj}$ channel events are required not to contain another electron, muon, or $\tauh$ candidate.
The $0\ell \mathrm{jj}$ channel selects events without a well-identified electron,
muon, or $\tauh$ candidate. The veto on additional leptons maintains high efficiency for compressed mass spectra scenarios and simultaneously reduces the SM backgrounds.
To further suppress QCD multijet background events containing large \ptmiss from jet
mismeasurements, the minimum azimuthal separation between any jet with $\pt > 30\GeV$ and the direction of the missing transverse momentum vector is required to be greater than 0.5
($\abs{\Delta\phi_{\text{min}}(\ptvecmiss, j)} > 0.5$).
Muon, electron, and $\tauh$ candidates must have $8 < \pt < 40\GeV$, $10 < \pt < 40\GeV$, and $20 < \pt < 40\GeV$,
respectively. The upper bound on lepton \pt suppresses the $\PZ \to\ell\ell$ and $\PW \to\ell\nu_{\ell}$ backgrounds where the average $\pt(\ell)$ is about $m_{\PZ} / 2$ and $m_{\PW} / 2$, respectively.
The lower bound on $\tauh$ \pt is larger because of known difficulties reconstructing lower-\pt $\tauh$ candidates, namely that they do not produce a
narrow jet in the detector, which makes them difficult to distinguish from quark or gluon jets.
All leptons are required to have $\abs{\eta} < 2.1$ in order to
select high quality and well-isolated leptons within the tracker acceptance. This requirement is 99\% efficient for signal events.
Lepton candidates are also required to pass
the reconstruction, identification, and isolation criteria described in Section~\ref{sec:obj}.

In addition to the $0\ell$ or $1\ell$ selection, the following requirements are imposed. The event is required to have $\ptmiss > 250\GeV$, which largely suppresses the $\PZ \to\ell\ell$ and QCD multijet backgrounds.
In order to reduce top quark pair contamination, the event is required not to have any jet identified as a \cPqb\ quark jet, following the description in Section 3;
only jets with $\pt > 30\GeV$, $\abs{\eta} < 2.4$, and separated from the leptons by $\DR > 0.3$ are considered for \PQb tags.
In the $1\ell$ channels, a minimum threshold on the transverse mass between the lepton and the
\ptmiss is imposed to minimize backgrounds with {\PW}  bosons. It is required that $\mT(\ell,\ptmiss) > 110\GeV$, \ie, beyond the Jacobian $m_{\PW}$ peak. The lepton- and \ptmiss-based
requirements described in this paragraph will be referred to as the ``central selection.''

The VBF signal topology is characterized by the presence of two jets in the forward
direction, in opposite hemispheres, and with large dijet invariant mass~\cite{VBFSbottom,VBFStop,VBFSlepton,VBFZprime,VBFHN,VBFY2,VBFHiggsInv1,VBFHiggsInv2}.
On the other hand, the jets in background events are mostly central and have small dijet invariant masses.
Additionally, the outgoing partons in VBF signal processes must carry relatively large \pt since they must have
enough energy (and be within the detector acceptance) to produce a pair of heavy SUSY particles (as shown in Fig.~\ref{fig:feynVBF}). Therefore, the ``VBF selection''
is imposed by requiring at least two jets with $\pt > 60\GeV$ and $\abs{\eta} < 5.0$. In the $1\ell \mathrm{jj}$ channels, only jets separated from the leptons by $\DR > 0.3$ are considered.
All pairs of jet candidates passing the above requirements and having $\abs{\Delta\eta} > 3.8$ and $\eta_{1}\eta_{2} < 0$ are combined to form VBF dijet candidates.
In the rare cases ($<$1\%) where selected events contain more than one dijet candidate satisfying the VBF criteria, the VBF dijet candidate with the largest
dijet mass is chosen since it is 97\% likely to result in the correct VBF dijet pair for signal events.
Selected dijet candidates are required to have $\mjj > 1\TeV$.

The signal region (SR) is defined as the events that satisfy the central and VBF selection criteria.

\section{Background estimation}
\label{sec:bkg}

The general methodology used for the estimation of background contributions in the SR is similar for all search channels and is based on both simulation and data.
Background-enriched control regions (CR) are constructed by applying selections orthogonal to those for the SR.
These CRs are used to measure the efficiencies of the VBF and central selections (the probability for a background component to satisfy the VBF and central selection criteria),
determine the correction factors to account for these efficiencies, and derive the shapes of the \mT and \mjj background distributions in the SR.
The correction factors are determined by assessing the level of agreement in the yields between data and simulation.
The shapes of distributions are derived directly from the
data in the CR, whenever possible, or
from the MC simulated samples when correct modeling by simulation is validated in the dedicated CRs.
For each final state, the same trigger is used for the CRs as for the corresponding SR.

The production of \ttbar events
represents the largest background source in the
$\Pe \mathrm{jj}$ and $\mu \mathrm{jj}$ channels (approximately 57--64\% of the total background), and the second largest
background source for the $\tauh \mathrm{jj}$ channel (approximately 29\% of the total background).
In the $0\ell \mathrm{jj}$ final state, since the combination of the lepton
and \PQb jet vetoes reduces this background to only approximately 5\% of the total background rate, its contribution is determined entirely from simulation.
The \ttbar background yields in the $1\ell \mathrm{jj}$ channels are evaluated using the following equation:
\begin{linenomath*}
\begin{equation}
N_{\ttbar}^{\text{pred}} = N_{\ttbar}^{\mathrm{MC}} \, SF_{\ttbar}^{\mathrm{CR}},
\label{eq:bgsr}
\end{equation}
\end{linenomath*}
where $N_{\ttbar}^{\text{pred}}$ is the predicted \ttbar background yield in the SR,
$N_{\ttbar}^{\mathrm{MC}}$ is the \ttbar rate predicted by simulation for the SR selection,
and $SF_{\ttbar}^{\mathrm{CR}}$ is the data-over-simulation correction factor, given by the ratio of observed data events to
the \ttbar yield in simulation, measured in a \ttbar enriched CR.
The numerator in the calculation of each correction factor is estimated by subtracting from data the contribution from other background events different from that under study,
and the statistical uncertainty is propagated to the $SF_{\ttbar}^{\mathrm{CR}}$ uncertainty.

The event selection criteria used to define the \ttbar
CR must not bias the correction factor $SF_{\ttbar}^{\mathrm{CR}}$. The simulated samples are used to check the closure of
this method, ensuring that the lepton kinematics, the composition of the events, and the \mT and \mjj shapes are similar between the CRs and the SR.
The closure tests demonstrate that the background determination techniques, described in detail below, reproduce the expected background distributions in both rate and shape to within the
statistical uncertainties. Various control samples are also utilized to validate the correct determination of the correction factors with the data.

\begin{figure}
  \centering

    \includegraphics[width=0.45\textwidth]{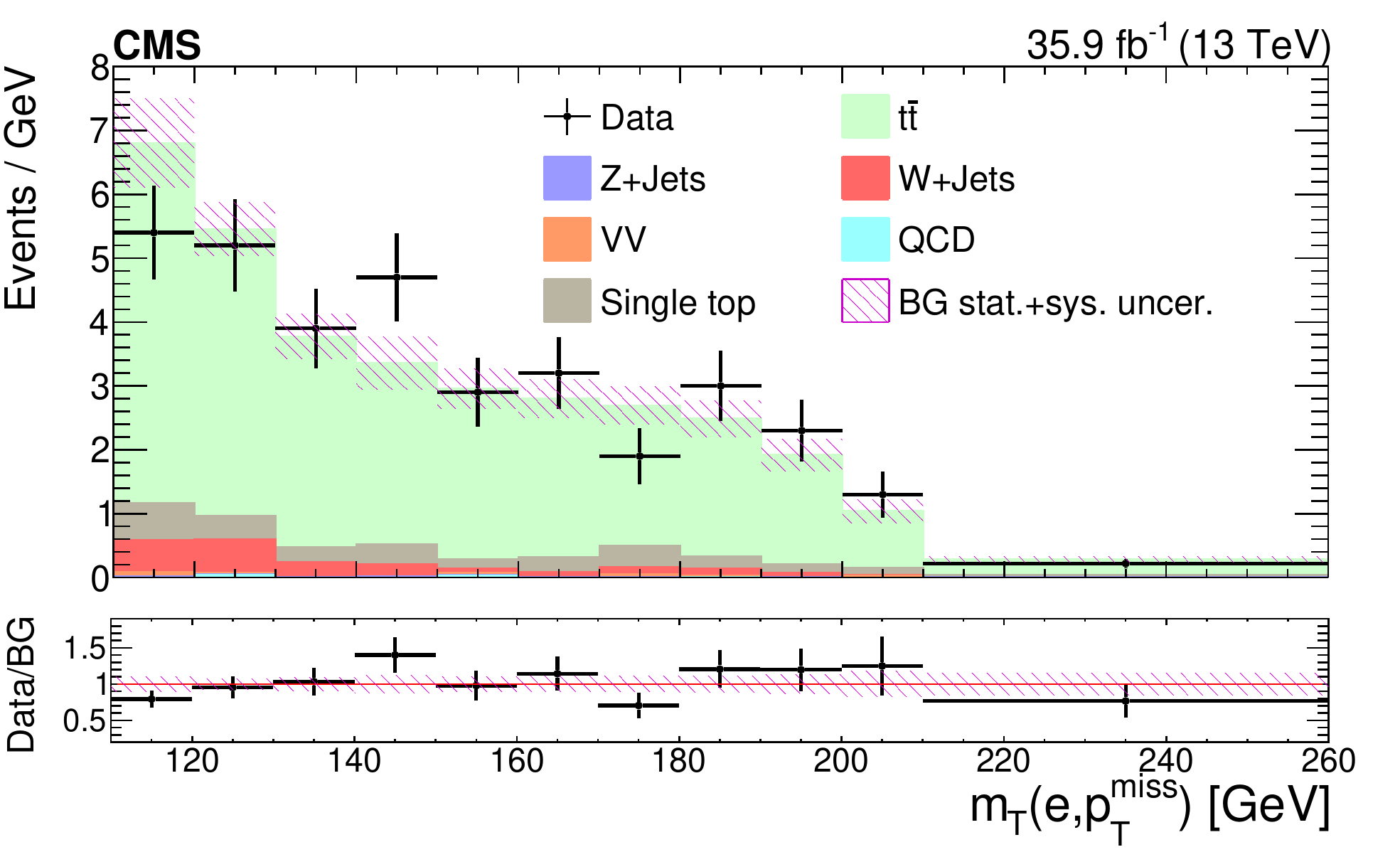}
    \includegraphics[width=0.45\textwidth]{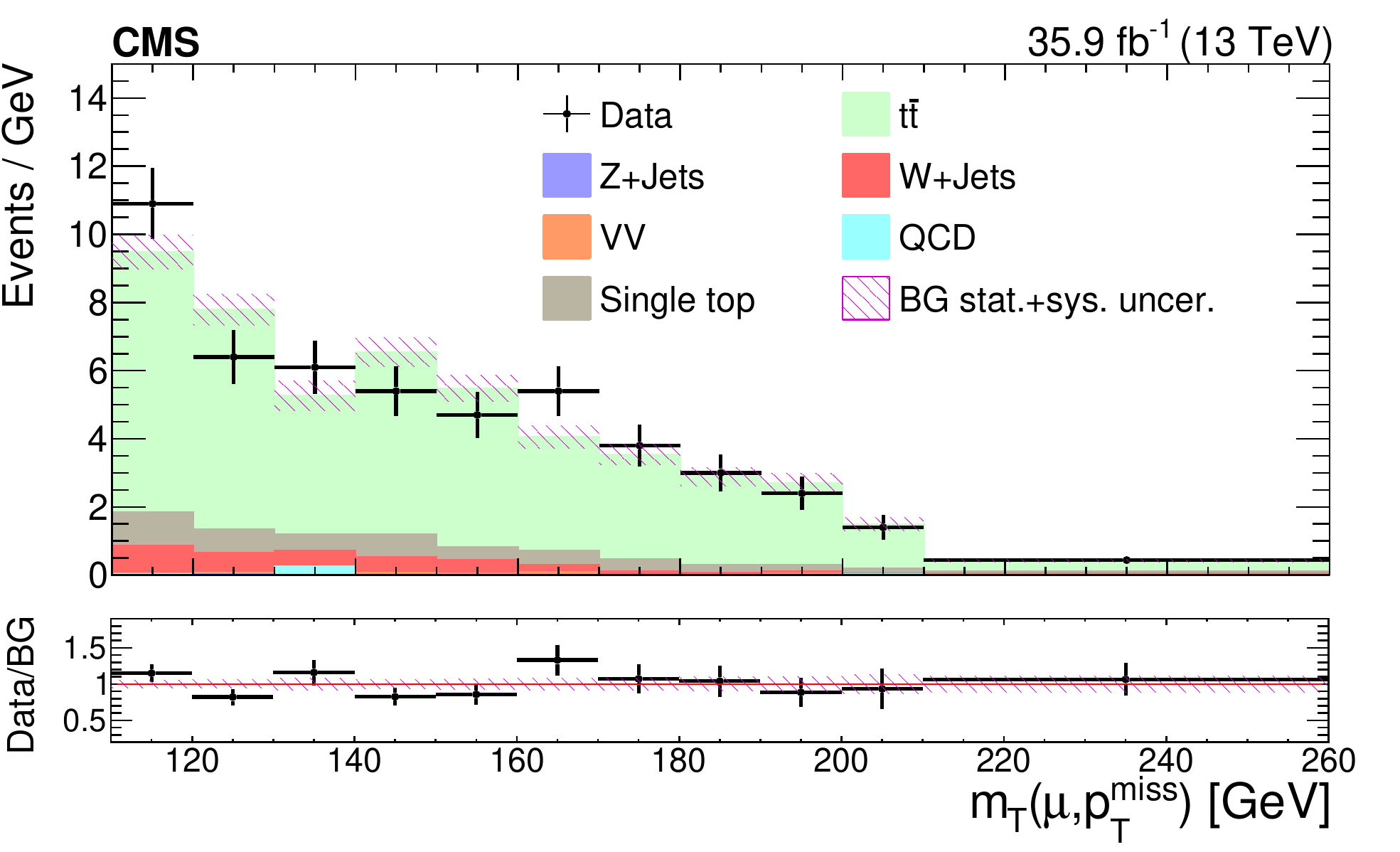}\\
    \includegraphics[width=0.45\textwidth]{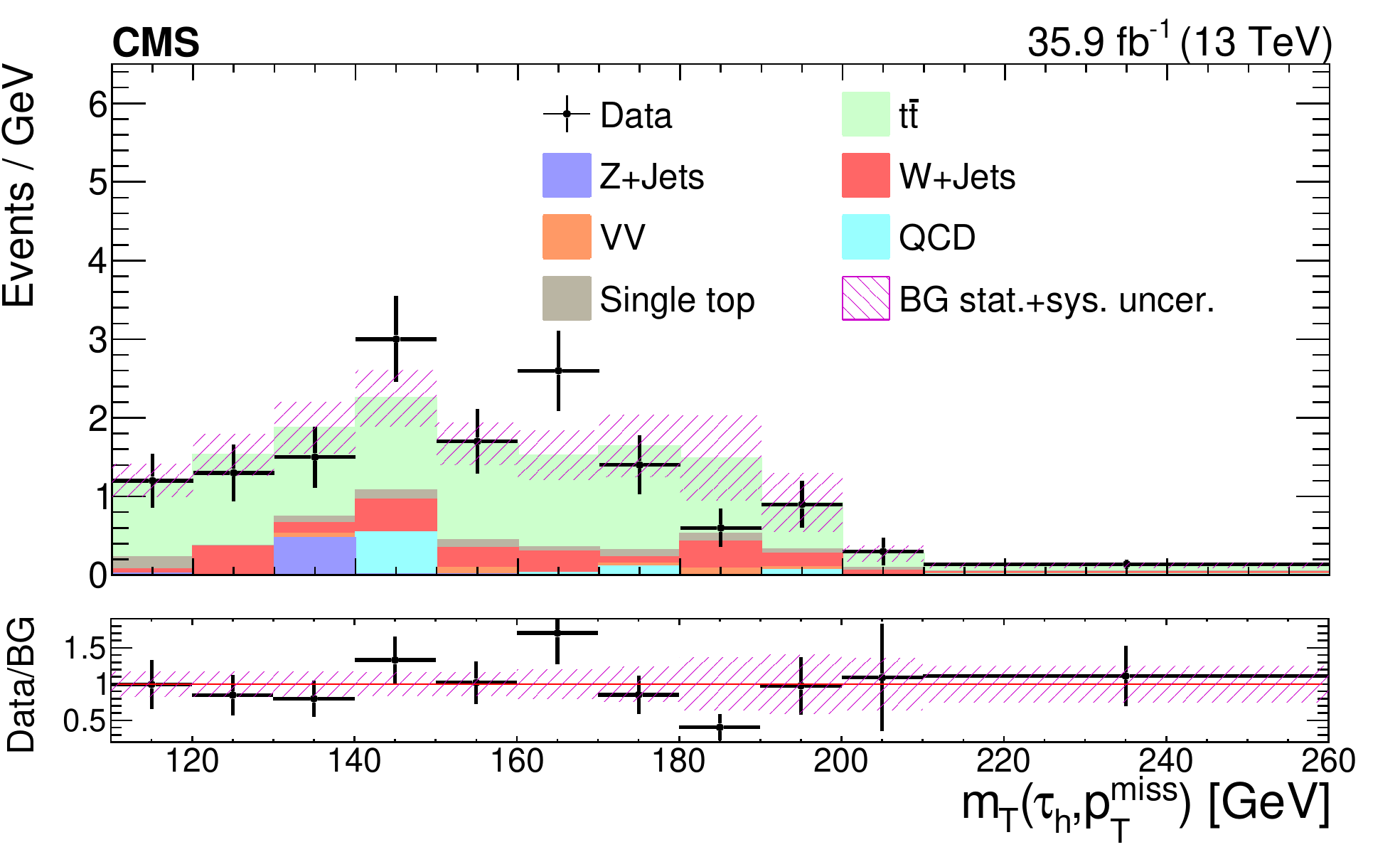}
    \includegraphics[width=0.45\textwidth]{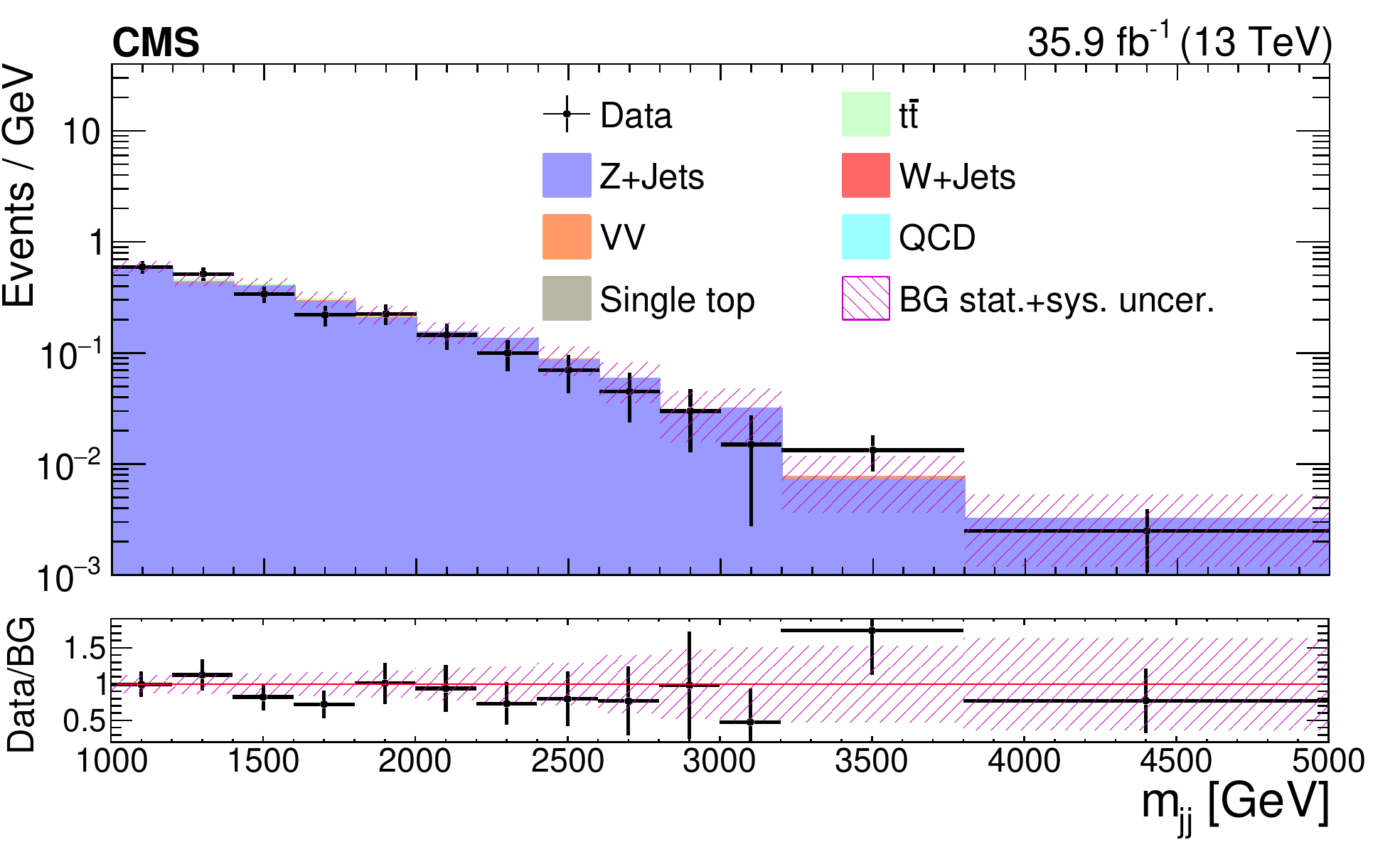}

  \caption{The $\mT$ distributions in the \ttbar control regions: (upper left) CR1$_{\Pe}$, (upper right) CR1$_{\Pgm}$, and (lower left) CR1$_{\tauh}$; (lower right) the \mjj distribution for $\PZ
(\to \PGn\PAGn)$+jets CR2 of the $0\ell \mathrm{jj}$ channel.}
  \label{fig:TTBarRegionInvertVBF}
\end{figure}

The \ttbar CR is obtained with similar selections to the SR, except
requiring one jet tagged as a \cPqb\ quark jet. These control samples with 1 \PQb-tagged jet
are referred to as CR$_{\Pe}$, CR$_{\Pgm}$, and CR$_{\tauh}$.
The 1 \PQb-tagged jet requirement significantly increases the \ttbar purity of the control samples while still
ensuring that those control samples contain the same kinematics and composition of misidentified leptons as the SR.
The \ttbar purity of the resulting data CR, determined from simulation, depends on the final state, ranging from 67 to
83\%. The measured data-over-simulation correction factors $SF_{\ttbar}^{\mathrm{CR}}$ are $0.8 \pm 0.3$, $0.8 \pm 0.2$, and
$1.3 \pm 0.5$ for the $\Pe \mathrm{jj}$, $\Pgm \mathrm{jj}$, and $\tauh \mathrm{jj}$ channels, respectively. The quoted uncertainties are based on the statistics in data and the simulated samples. Systematic uncertainties are discussed in Section~\ref{sec:sys}.
Figure~\ref{fig:TTBarRegionInvertVBF} contains the $\mT$ distributions for the \ttbar control regions: (upper left) CR$_{\Pe}$, (upper right) CR$_{\Pgm}$, and (lower left) CR$_{\tauh}$.
The correction factors $SF_{\ttbar}^{\mathrm{CR}}$ have been applied to the MC simulation distributions shown in Fig.~\ref{fig:TTBarRegionInvertVBF}.
The \mT shapes between data and simulation are consistent within statistical uncertainties (the bands in the data over background (BG) ratio distributions represent the statistical uncertainties of the data and simulated samples). Therefore, the $\ttbar$ $\mT$ shapes in the SR are taken directly from simulation.

In general, the {\PW}+jets and {\PZ}+jets backgrounds represent an important contribution in the $0\ell \mathrm{jj}$ and $1\ell \mathrm{jj}$ channels, and their
contributions to the SR are evaluated using two control regions CR1 and CR2 (defined below for each BG component) and using the equation:
\begin{linenomath*}
\begin{equation}
N_{\mathrm{BG}}^{\text{pred}} = N_{\mathrm{BG}}^{\mathrm{MC}} \, SF_{\mathrm{BG}}^{\mathrm{CR1}}(\text{central}) \,
SF_{\mathrm{BG}}^{\mathrm{CR2}}(\text{VBF}),
\label{eq:bgsr2}
\end{equation}
\end{linenomath*}
where $N_{\mathrm{BG}}^{\text{pred}}$ is the predicted BG yield in the SR,
$N_{\mathrm{BG}}^{\mathrm{MC}}$ is the rate predicted by simulation (with BG = {\PW}+jets and {\PZ}+jets) for the SR selection,
$SF_{\mathrm{BG}}^{\mathrm{CR1}}(\text{central})$ is the data-over-simulation correction factor for the central selection, given by the ratio of data to the BG simulation in control region CR1,
and $SF_{\mathrm{BG}}^{\mathrm{CR2}}(\text{VBF})$ the data-over-simulation correction factor for the efficiency of the VBF selections as determined in another background enriched control sample CR2.

The production of $\PZ (\to \PGn\PAGn)$+jets is the main SM background to the $0\ell \mathrm{jj}$ SR,
with a similar signal topology from the neutrino contributions to \ptmiss, and
is therefore mostly irreducible. The strategy for the $\PZ (\to \PGn\PAGn)$+jets background estimation is to use simulation to model the \ptmiss distribution, and jet and lepton vetoes.
The background yields predicted by the MC simulated samples are corrected for observed differences with respect to the data in the CRs, and scaled to the fraction of events
passing the VBF selection, derived from data. The modeling of the \mjj distribution is checked in the CRs.
Two CRs are used to verify the MC simulation, estimate acceptance corrections used to scale the MC simulation yields, and measure the fraction of events passing the VBF selection.
The control regions are defined by treating muons as neutrinos in the $\PZ \to \MM$ decay mode.
The first control region (CR1$_{\PZ}$) is a $\PZ (\to \MM)$+two jets sample used to validate modeling of geometric and kinematic acceptance of leptons. 
The invariant mass of the opposite-sign dimuon system must be consistent with the $\PZ$-boson mass ($60$-$120\GeV$). 
The two muons are treated as neutrinos, excluding the muon \pt vectors from \ptvecmiss, and require $\ptmiss> 250\GeV$ together with a veto on \PQb-tagged jets and additional leptons, as in the SR.
The measured data-over-simulation correction factor is $0.95 \pm 0.02\stat$.
Adding the VBF selection defines CR2$_{\PZ}$. The {\PZ}+jets prediction from simulation in CR2$_{\PZ}$ is corrected with the measured data-over-simulation correction factor from CR1$_{\PZ}$ to ensure
$SF_{\mathrm{BG}}^{\mathrm{CR2}}$ represents a correction for the efficiency of the VBF selection (correlations between the uncertainties of CR2$_{\PZ}$ and CR1$_{\PZ}$ are also taken into account). 
The ratio of CR2$_{\PZ}$ to CR1$_{\PZ}$ events in the data gives the fraction of $\PZ (\to \PGn\PAGn)$+jets events passing the VBF topology selection.
The measured data-over-simulation correction factor in CR2$_{\PZ}$ is $0.92 \pm 0.12\stat$.
Figure~\ref{fig:TTBarRegionInvertVBF} (lower right) shows the \mjj distribution in $\PZ (\to \PGn\PAGn)$+jets CR2$_{\PZ}$, which shows agreement between the data and the corrected {\PZ}+jets 
prediction from simulation.

The production of {\PW}+jets events presents another important source of background for all the search channels.
For the $1\ell \mathrm{jj}$ channels, control samples enriched in {\PW}+jets events, with about 65\% purity according to simulation, are obtained by
requiring similar criteria to the SR, except with an inverted VBF selection (failing the VBF selection as defined in Section 5). The inverted VBF selection enhances the
{\PW}+jets background yield by two orders of magnitude, while suppressing the VBF signal contamination to negligible levels. This control region, CR1$_{\PW}$, is used to obtain a correction factor 
for the efficiency of the central selection, $SF_{\PW+\text{jets}}^{\mathrm{CR1}}(\text{central})$. This correction factor is determined to be $0.97 \pm
0.10$ and $1.10 \pm 0.10$, for the $\Pe \mathrm{jj}$ and $\Pgm \mathrm{jj}$ channels, respectively.
The quoted uncertainties are based on the statistics in data and the simulated samples.
For the $\tauh \mathrm{jj}$ channel, it is difficult to obtain
a control sample enriched in {\PW}+jets events because there is a significant contribution from QCD multijet events.
Therefore, the average of the correction factors obtained for the $\Pe \mathrm{jj}$ and $\Pgm \mathrm{jj}$ channels, $1.04 \pm 0.13$, is used to scale the {\PW}+jets prediction from simulation in the
$\tauh \mathrm{jj}$ channel. This approach is justified since the $\PW(\to\tau\Pgngt)$+jets prediction from simulation is corrected to account for slight differences in the $\tauh$ identification efficiency observed in
data. This is further supported by the fact that the modeling of the
VBF efficiency at simulation level is uncorrelated with the decay of the {\PW} boson.
The relatively small difference in mass between {\PW} and {\PZ} bosons (compared to the energy scale of the SR), which allows the use of a control sample (CR2$_{\PW}$)
enriched with {\PZ}+jets events to measure the VBF selection efficiency for the {\PW}+jets background in the $1\ell \mathrm{jj}$ channels. This second control sample is obtained by selecting events containing two
muons with $\pt > 30\GeV$, treating only one muon as a neutrino to recalculate $\ptvecmiss$, and otherwise similar selections to the SR.
Since the efficiency and momentum scale of muons are known at the 1--2\% level, any disagreement between data and simulation
in this $\PZ (\to \MM)$+jets control sample is used to measure the correction factor for the modeling of the VBF selection efficiency in {\PW}+jets events.
The correction factor $SF_{\PW+\mathrm{jets}}^{\mathrm{CR2}}(\text{VBF})$ determined from the CR2$_{\PW}$ control sample is measured to be $1.18 \pm 0.09$ (correlations between the uncertainties of 
$SF_{\PW+\mathrm{jets}}^{\mathrm{CR2}}$ and $SF_{\PZ+\mathrm{jets}}^{\mathrm{CR2}}$ are taken into account). 
To validate the correction factors, the {\PW}+jets rate in samples with $\mT < 110\GeV$ is scaled
by $SF_{\PW+\mathrm{jets}}^{\mathrm{CR1}}(\text{central})$ and $SF_{\PW+\mathrm{jets}}^{\mathrm{CR2}}(\text{VBF})$, and agreement between the data and the corrected {\PW}+jets prediction from simulation
is observed.

In the $0\ell \mathrm{jj}$ channel, $\PW (\to \ell \nu_{\ell})$+jets events can enter the SR, because of the contribution to \ptmiss from the neutrino, if the accompanying charged lepton
fails the lepton veto criteria. To determine the contribution of $\PW (\to \ell \nu_{\ell})$+jets background to the
$0\ell \mathrm{jj}$ SR, a similar procedure to the $\PZ (\to \PGn\PAGn)$+jets background estimation methodology is used. The muon veto is replaced with a one-muon requirement to obtain a $\PW (\to
\mu \Pgngm)$ plus two jets sample, requiring $60 < m_{\textrm{T}}(\mu,\ptmiss) < 100\GeV$, treating the muon as undetected, and requiring $\ptmiss > 250\GeV$ as in the SR selection.
The simulated samples are used to demonstrate that substituting the muon veto for a one-muon requirement does not affect the shapes of the \ptmiss and VBF jet kinematic distributions.
The measured data-over-simulation correction factor is $0.90 \pm 0.02\stat$. 
The control region is obtained by adding the VBF topology selection, and has a measured data-over-simulation correction factor of $0.90 \pm 0.08\stat$.

The QCD multijet background is only important in the $0\ell \mathrm{jj}$ and $\tauh \mathrm{jj}$ channels.
Among the main discriminating variables against QCD multijet events are the VBF selection criteria,
the minimum separation between \ptvecmiss and any jet $\abs{\Delta\phi_{\text{min}}(\ptvecmiss, j)}$, and $\tauh$ isolation.
Thus, the QCD multijet background estimation methodology utilizes CRs obtained by inverting
these requirements. In the $\tauh \mathrm{jj}$ channel, the QCD multijet background is estimated using a completely data-driven approach which relies on the matrix (``ABCD'') method.
The regions are defined as follows:

\begin{itemize}
  \item CR$A$: inverted VBF selection; pass the nominal (tight) $\tauh$ isolation;
  \item CR$B$: inverted VBF selection; fail the nominal $\tauh$ isolation but pass loose $\tauh$ isolation;
  \item CR$C$: pass the VBF selection; fail the nominal $\tauh$ isolation but pass loose $\tauh$ isolation and;
  \item CR$D$: pass the VBF selection; pass the nominal $\tauh$ isolation
\end{itemize}

The QCD multijet component $N_{\text{QCD}}^{\mathrm{i}}$ in regions $i=\mathrm{CR}A,\mathrm{CR}B,\mathrm{CR}C$ is estimated by subtracting non-QCD backgrounds (predicted using simulation)
from data ($N_{\text{QCD}}^{\mathrm{i}}=N_{\text{Data}}^{\mathrm{i}}-N_{\neq \text{QCD}}^{\mathrm{i}}$). The QCD multijet component in CR$D$ (\ie, the SR) is then estimated to
be $N_{\text{QCD}}^{SR} = N_{\text{QCD}}^{\mathrm{CRA}} \, N_{\text{QCD}}^{\mathrm{CRC}} / N_{\text{QCD}}^{\mathrm{CRB}}$, where
$N_{\text{QCD}}^{\mathrm{CRC}} / N_{\text{QCD}}^{\mathrm{CRB}}$ is referred to as the ``pass-to-fail VBF'' transfer factor ($TF_{\text{VBF}}$). Said differently, the yield of QCD
multijet events in data with an inverted VBF selection is extrapolated to the SR using the transfer factor $TF_{\text{VBF}}$, which is measured in
data samples enriched with QCD multijet events that fail the nominal $\tauh$ isolation criteria but satisfy the loose $\tauh$ isolation working point (henceforth referred to as ``inverted
$\tauh$ isolation'' or ``nonisolated $\tauh$'').
The purity of the QCD multijet events is approximately 53--77\% depending on the CR.
The shape of the $\mT(\tauh,\ptmiss)$ distribution is obtained from CR$B$ (from the nonisolated $\tauh$ plus inverted VBF control
sample). This ``ABCD'' method relies on $TF_{\text{VBF}}$ being unbiased by the $\tauh$ isolation requirement. A closure test of this assumption is provided using the
simulated QCD multijet samples, resulting in agreement at a 5\% level and within the statistical uncertainties.

In the $0\ell \mathrm{jj}$ channel, the contribution from QCD multijet production is estimated using the number of events passing the analysis selection except the
$\abs{\Delta \phi_{\text{min}} ( \ptvecmiss, j)}$ requirement.
The QCD multijet purity in this CR is about 74\% according to simulation.
The \mjj distribution of the non-QCD background is subtracted from the \mjj data distribution, and the resultant QCD multijet \mjj distribution from data is
scaled by the efficiency to inefficiency ratio of the
$\abs{\Delta \phi_{\text{min}}( \ptvecmiss, j )}$ requirement,
$TF_{\Delta\phi}$. The transfer factor $TF_{\Delta\phi} = 0.06 \pm 0.01$ is determined using the simulated QCD multijet samples and validated using data control samples
obtained by selecting events that fall in the dijet mass window $500 < \mjj < 1000\GeV$.

\section{Systematic uncertainties}
\label{sec:sys}

The main contributions to the total systematic uncertainty in the background predictions arise from the closure tests and
from the statistical uncertainties associated with the data CRs used to determine the
$SF^{\mathrm{CR1}}_{\mathrm{BG}}\text{(central)}$, $SF^{\mathrm{CR2}}_{\mathrm{BG}}\text{(VBF)}$, $TF_{\text{VBF}}$, and $TF_{\Delta\phi}$ factors.
The relative systematic uncertainties on the product $SF^{\mathrm{CR1}}_{\mathrm{BG}}\text{(central)}SF^{\mathrm{CR2}}_{\mathrm{BG}}\text{(VBF)}$ related to the statistical precision in the CRs
range between 8 and 42\%, depending on the background component and search channel. For $TF_{\text{VBF}}$ and $TF_{\Delta\phi}$, the statistical uncertainties
lie between 13 and 22\%. The systematic uncertainties in the $SF^{\mathrm{CR1}}_{\mathrm{BG}}\text{(central)}$, $SF^{\mathrm{CR2}}_{\mathrm{BG}}\text{(VBF)}$, $TF_{\text{VBF}}$, and
$TF_{\Delta\phi}$ factors, evaluated from the closure tests and cross-checks with data, range from 9 to 33\%, depending on the channel. 
Additionally, although the background \mT and \mjj shapes between data and simulation are consistent within statistical uncertainties, data/BG ratios of the 
\mT and \mjj distributions are fit with a first-order polynomial, and the deviation of the fit from unity, as a function of \mT or \mjj, is conservatively taken as the systematic uncertainty on the 
shape. This results in up to $\approx$10\% systematic uncertainty in a given \mT or \mjj bin.

Less significant contributions to the systematic uncertainties arise from contamination by non-targeted background sources to the CRs used to measure
$SF^{\mathrm{CR1}}_{\mathrm{BG}}\text{(central)}$ and $SF^{\mathrm{CR2}}_{\mathrm{BG}}\text{(VBF)}$, and
from the uncertainties in these correction factors caused by uncertainties in the lepton identification efficiency, lepton energy and momentum scales, \ptmiss scale, and trigger efficiency.

The efficiencies for the electron and muon reconstruction, identification, and isolation requirements are measured with the ``tag-and-probe"
method~\cite{MUONreco,Khachatryan:2015hwa} with a resulting uncertainty of $\le$2\%, dependent on \pt and $\eta$. The total efficiency for the $\tauh$ identification and isolation requirements
is measured from a fit to the $\PZ\to\tau\tau\to\mu\tauh$ visible mass distribution in a sample selected with one isolated muon trigger candidate with $\pt > 24\GeV$,
leading to a relative uncertainty of 5\% per $\tauh$ candidate~\cite{1748-0221-11-01-P01019}.
The \ptmiss scale uncertainties contribute via the jet energy scale (2--5\% depending on $\eta$ and \pt) and unclustered energy scale
(10\%) uncertainties, where ``unclustered energy" refers to energy from a reconstructed object that is not assigned to a jet with $\pt >10\GeV$ or to a lepton with $\pt >10\GeV$.
A \ptmiss-dependent uncertainty in the measured trigger efficiency results in a 3\% uncertainty in the signal and background predictions that rely on simulation. The trigger efficiency is measured
by calculating the fraction of {\PW}+jets events (selected with the same single-$\Pgm$ trigger), that also pass the same trigger that is used to define the SR.

The signal and minor backgrounds, estimated solely from simulation, are
affected by similar sources of systematic uncertainty.
For example, the uncertainties in the lepton identification efficiency, lepton energy and momentum scale, \ptmiss scale,
trigger efficiency, and integrated luminosity uncertainty of 2.5\%~\cite{CMS-PAS-LUM-17-001} also contribute to the systematic uncertainty in the signal.

The signal event acceptance for the VBF selection depends on the reconstruction and identification
efficiency and jet energy scale of forward jets.
The total efficiency for the jet reconstruction and identification requirements is $>$98\% for the
entire $\eta$ and \pt range, as validated through the agreement observed between data and simulation in the
$\eta$ distribution of jets, in particular at high $\eta$, in CRs enriched with \ttbar
background events. Among the dominant uncertainties in the signal acceptance is the modeling of the kinematic properties of jets, and thus the efficiency to select VBF topologies for forward jets in
the \MADGRAPH simulation. This is investigated by comparing the predicted and measured \mjj  spectra in the {\PZ}+jets CRs.
The level of agreement between the predicted and observed \mjj spectra is better than 9\%, which is assigned
as a systematic uncertainty in the VBF efficiency for signal samples.
The dominant uncertainty in the signal acceptance arises from the partial mistiming of signals in the forward region of the ECAL endcaps,
which led to a reduction in the L1 trigger efficiency. A correction for this effect was determined using an unbiased data sample.
This correction was found to be about 8\% for \mjj of 1\TeV and increases to about 19\% for \mjj greater than 3.5\TeV.
The uncertainty in the signal acceptance from the PDF set used in simulation is evaluated in accordance with the PDF4LHC recommendations~\cite{PDF4LHCRunII} by
comparing the results obtained using the
CTEQ6.6L, MSTW08, and NNPDF10  PDF sets~\cite{Nadolsky:2008zw,mrst2006,nnpdf10} with those from the default PDF set. 
It should be noted that the combined uncertainty on the signal yields and \mjj/\mT shapes due to scale variations on renormalization, factorization, 
and jet matching is found to be about 2\%, which is small compared to our estimate of 9\% using the {\PZ}+jets CRs. 
Other dominant uncertainties that contribute to the \mjj and \mT shape variations include the \ptmiss energy scale, \tauh energy scale, and jet energy
scale uncertainties.

\section{Results and interpretation}
\label{sec:res}

\begin{figure}[tbh!]
  \centering
     \includegraphics[width=0.45\textwidth]{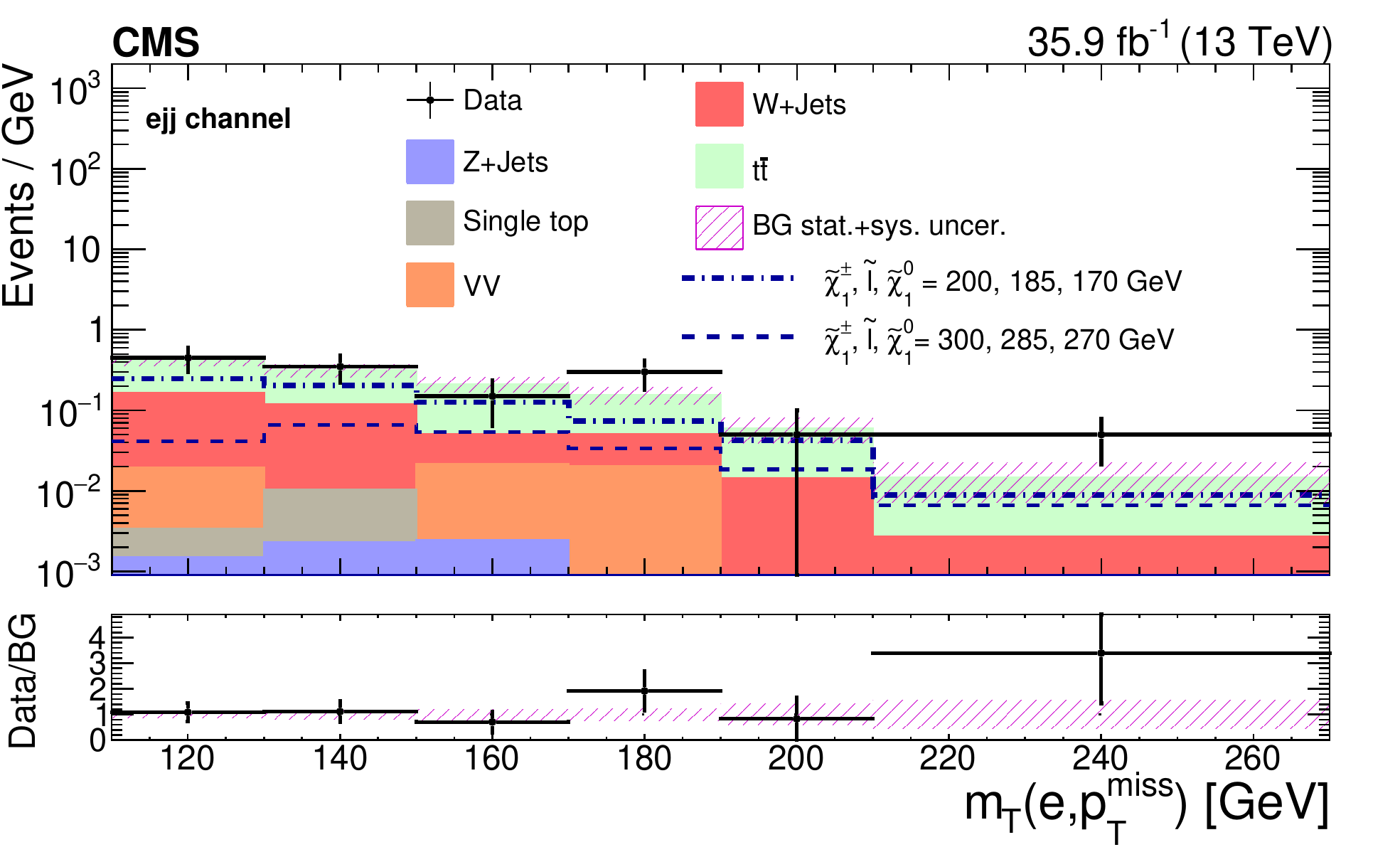}
     \includegraphics[width=0.45\textwidth]{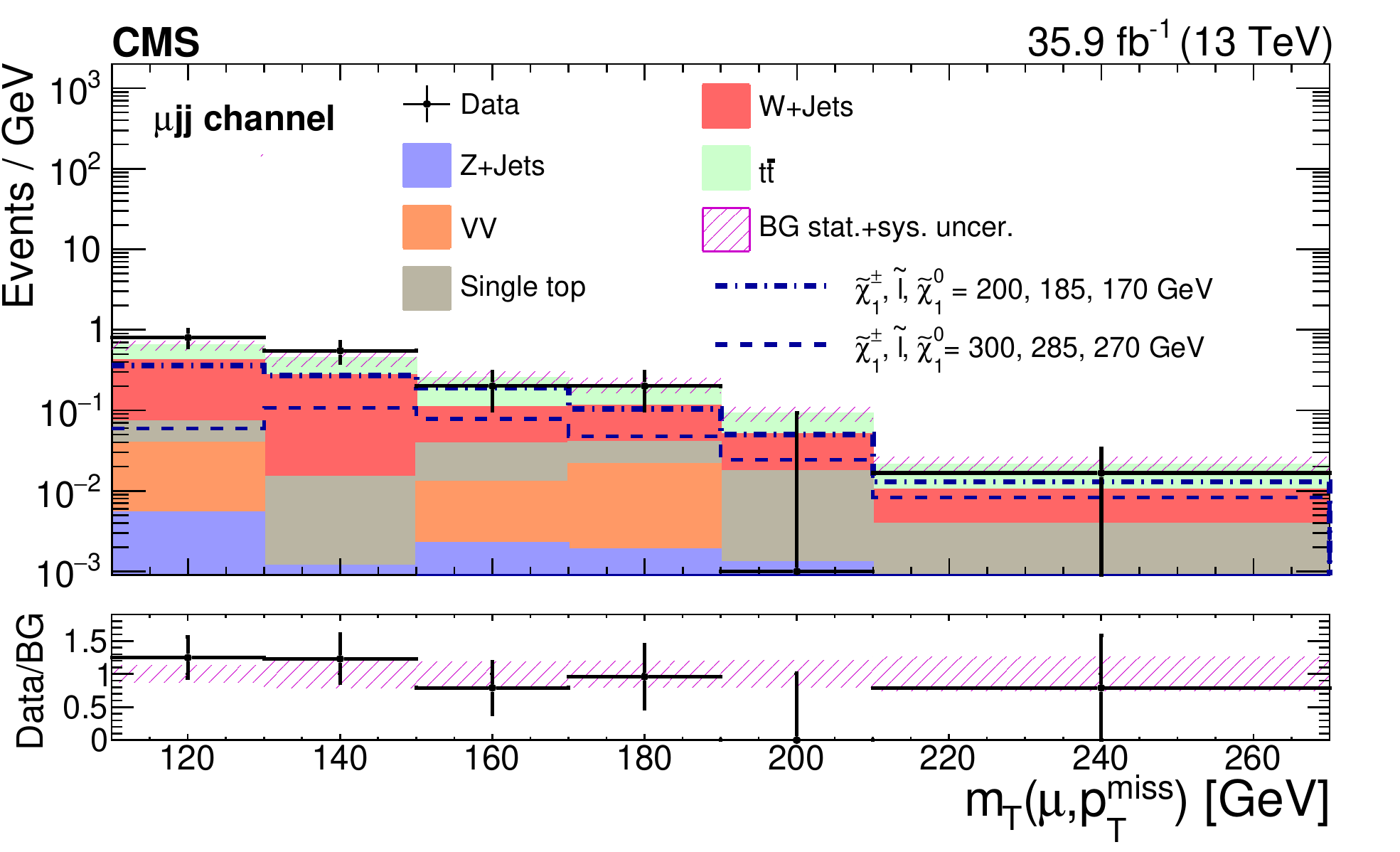}\\
     \includegraphics[width=0.45\textwidth]{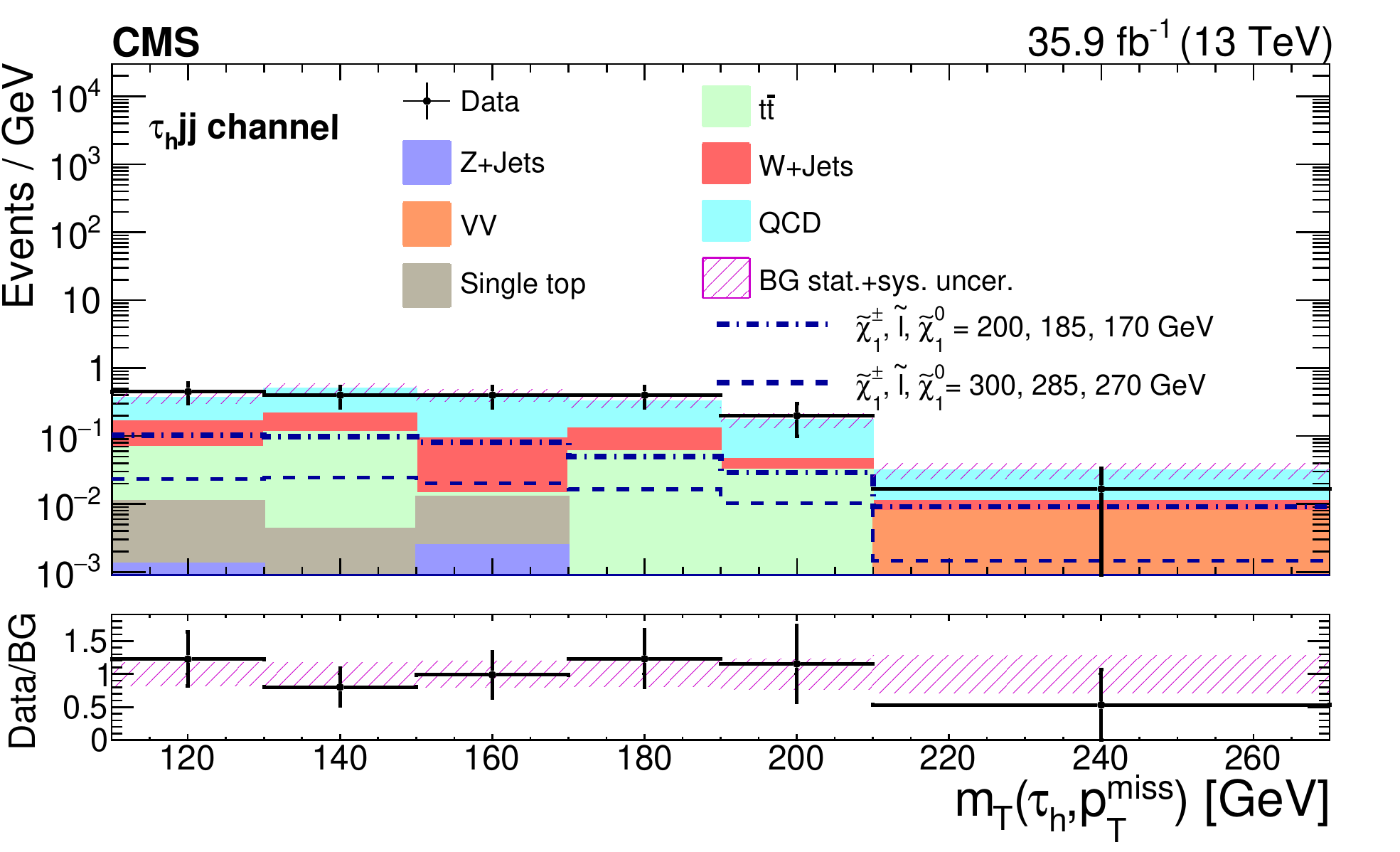}
     \includegraphics[width=0.45\textwidth]{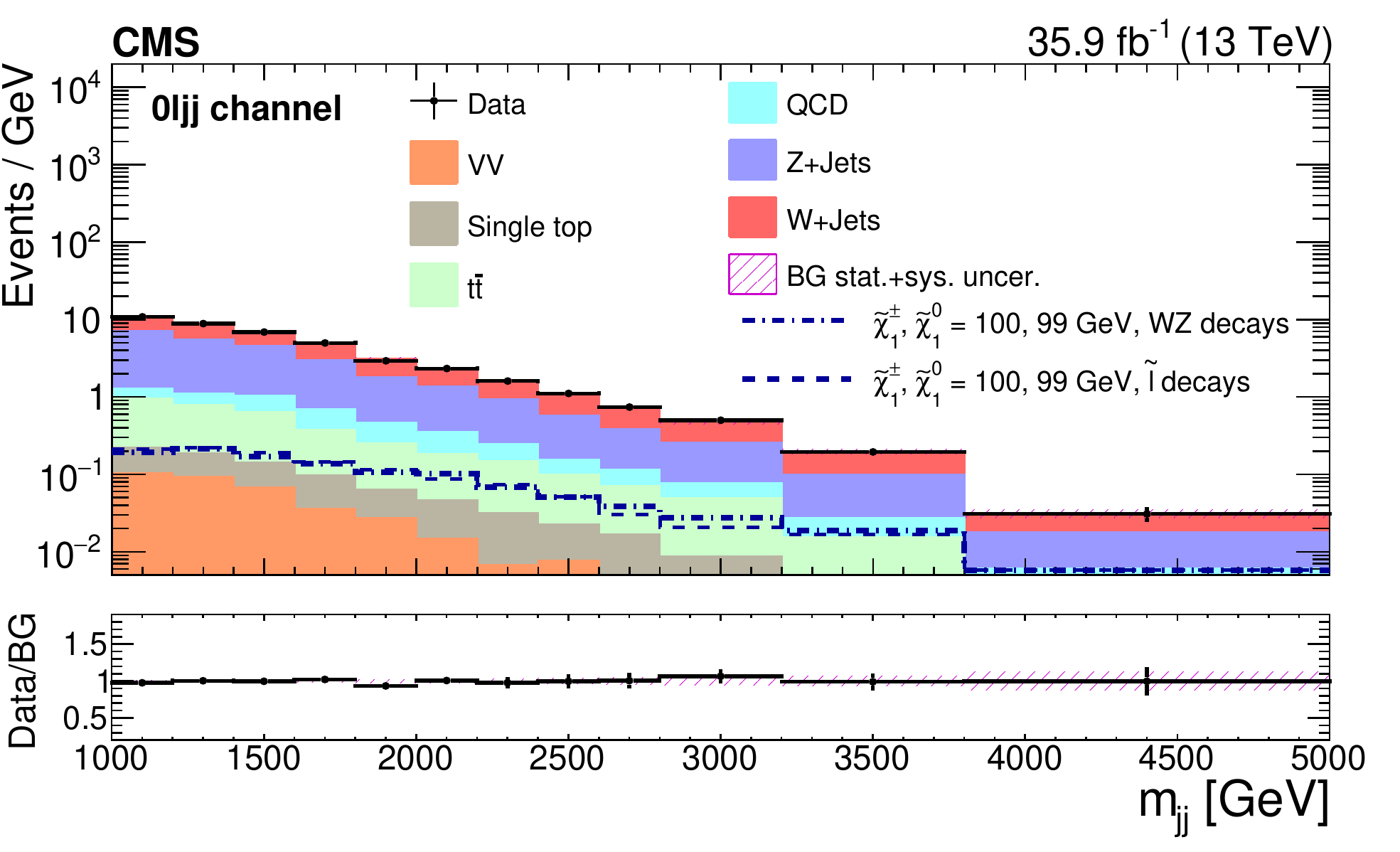}
  \caption{The observed \mT and \mjj distributions in the
$\Pe \mathrm{jj}$ (upper left), $\mu \mathrm{jj}$ (upper right), $\tauh \mathrm{jj}$ (lower left), and $0\ell \mathrm{jj}$ (lower right) signal regions compared with the post-fit SM background yields from
the fit described in the text. The pre-fit background yields and shapes are determined using data-driven methods for the major backgrounds, and based on simulation for the smaller backgrounds. Expected signal distributions are
overlaid. The last bin in the \mT distributions of the $1\ell \mathrm{jj}$ channels include all events with $\mT > 210\GeV$. The last bin of the \mjj
distributions of the
$0\ell \mathrm{jj}$ channel include all events with $\mjj > 3800\GeV$.}
  \label{fig:SRPlots}
\end{figure}

Table 1 lists the number of observed events in data as well as the predicted background contributions in the SR for each channel, integrating over \mjj and \mT bins.
Figure~\ref{fig:SRPlots} shows the predicted SM background, expected signal, and observed data rates in bins of \mT for the $1\ell \mathrm{jj}$ channels and
bins of \mjj in the $0\ell \mathrm{jj}$ channel. The bin sizes in the distributions of Fig.~\ref{fig:SRPlots} are chosen to maximize the signal significance of the analysis.
No significant excess of events is observed above the SM prediction in any of the search regions.
Therefore the search does not reveal any evidence for new physics.

\begin{table*}[htbp]
  \centering{
  \topcaption{The number of observed events and corresponding pre-fit background predictions, where ``pre-fit'' refers to the predictions determined as described in the text, before constraints from 
the fitting procedure have been applied. The uncertainties include the statistical and systematic components.}
  \begin{tabular}{ l  c  c  c  c }\hline
    Process  & $\Pgm \mathrm{jj}$ & $\Pe \mathrm{jj}$ & $\tauh \mathrm{jj}$ & $0\ell \mathrm{jj}$  \\ \hline
    DY+jets         & $0.20 \pm 0.07$ & $0.10 \pm 0.04$    & $0.10 \pm 0.04$  & $3714 \pm 760$\\
    {\PW}+jets        & $13 \pm 3$     & $6 \pm 1$    & $7 \pm 2$  & $2999 \pm 620$\\
    VV              & $1.7 \pm 0.7$          & $1.5 \pm 0.6$  & $0.9 \pm 0.9$  & $77 \pm 18$\\
    \ttbar          & $13 \pm 4$          & $11 \pm 4$  & $5 \pm 3$  & $577 \pm 128$\\
    Single top  quark     & $2.2 \pm 0.9$        & $0.2 \pm 0.1$                     & $0.6 \pm 0.3$             & $104 \pm 10$\\
    QCD             & $0^{+0.2}_{-0}$        & $0^{+1.2}_{-0}$                     & $23 \pm 5$             & $546 \pm 69$\\ [\cmsTabSkip]
    Total BG        & $31 \pm 5$          & $19 \pm 5$  & $37 \pm 6$ & $8017 \pm 992$\\ [\cmsTabSkip]
    Data            &  36         & 29                    &  38            & 8408 \\
    \hline

  \end{tabular}
  }
  \label{table:SignalRegionTable}
\end{table*}

To illustrate the sensitivity of this search, the results are presented in the context of the $R$-parity  conserving MSSM and considering cases such as those
shown in Fig.~\ref{fig:feynVBF} for pure electroweak VBF production of charginos and neutralinos.
As mentioned previously, models with a bino-like \PSGczDo and wino-like \PSGczDt and \PSGcpmDo are considered. Since in this case the \PSGczDt and \PSGcpmDo belong to the same gauge
group multiplet, the \PSGczDt mass is set to $m_{\PSGczDt}=m_{\PSGcpmDo}$ and results are presented as a function of this common mass and mass difference $\Delta m \equiv m(\PSGczDt) -
m(\PSGczDo)$. Two scenarios have been considered:
$(i)$ the ``light slepton'' model where $\widetilde{\ell}$ is the next-to-lightest SUSY particle; and $(ii)$ the ``$\PW\PZ$'' model where sleptons are too heavy and thus \PSGcpmDo and
\PSGczDt decays proceed via $\PW^{*}$ and $\PZ^{*}$. The main difference between the two models is the branching ratio of \PSGcpmDo and \PSGczDt to leptonic final states.
It should be noted that the branching fractions to leptons are adapted to off-shell $\PW$ and $\PZ$ bosons. 
In the models shown in the top row of Fig.~\ref{fig:feynVBF}, the mass $m_{\widetilde{\ell}}$ of the intermediate slepton is parameterized in terms of a variable $x_{\widetilde{\ell}}$ as
\begin{linenomath*}
\begin{equation}
m_{\widetilde{\ell}} = m_{\PSGczDo} + x_{\widetilde{\ell}}(m_{\PSGcpmDo} - m_{\PSGczDo}),
\end{equation}
\end{linenomath*}
where $0 < x_{\widetilde{\ell}} < 1$. Results are presented for $x_{\widetilde{\ell}}=0.5$ in the ``$\widetilde{\ell}$-democratic" model where three sleptons ($m_{\widetilde{\ell}} =
m_{\PSe} =
m_{\PSGm} = m_{\PSgt}$) are light~\cite{CMSEWK}. The results are interpreted by assuming branching fractions
$\mathcal{B}(\PSGczDt\to \ell \widetilde{\ell} \to \ell\ell \PSGczDo)=1$ and $\mathcal{B}(\PSGcpmDo\to\nu_{\ell}\widetilde{\ell} \to\nu_{\ell}\ell\PSGczDo)=1$.
To highlight the evolution of the search sensitivity for compressed spectra with mass gap $\Delta m$, values between $\Delta m = 1$ and 50\GeV are studied for both the light slepton and
$\PW\PZ$ interpretations.
The signal selection efficiency for the $1\mu \mathrm{jj}$ ($1\Pe\mathrm{jj}$) channel in the light slepton model, assuming $\Delta m = 30$\GeV, is 0.9 (0.7)\% for
$m(\PSGcpmDo) = 100$\GeV and 2.5 (1.8)\% for $m(\PSGcpmDo) = 300$\GeV. Similarly, the signal selection efficiency for the $0\ell \mathrm{jj}$ channel,
assuming $\Delta m = 1$\GeV, is 2.8\% for $m(\PSGcpmDo) = 100\GeV$ and 5.3\% for $m(\PSGcpmDo) = 300\GeV$.

The calculation of the exclusion limit is obtained by using the \mT (\mjj) distribution in the $1\ell \mathrm{jj}$ ($0\ell \mathrm{jj}$)
to construct a combined profile likelihood ratio test statistic~\cite{Cowan:2010js} in bins of \mT (\mjj)  and computing a 95\% confidence level (\CL)
upper limit (UL) on the signal cross section using the asymptotic \CLs criterion \cite{Junk, CLs1, Cowan:2010js}.
Systematic uncertainties are taken into account as nuisance parameters, which are removed by profiling, assuming gamma function or
log-normal priors for normalization parameters, and Gaussian priors for mass spectrum shape uncertainties.
The combination of the four search channels requires simultaneous analysis of the
data from the individual channels, accounting for all statistical and systematic uncertainties
and their correlations. Correlations among backgrounds, both within a channel and across channels, are taken into consideration in the limit calculation.
For example, the uncertainty in the integrated luminosity is treated as fully correlated across channels. 
The uncertainties in the predicted signal yields resulting from the event acceptance variation with different sets of PDFs in a given \mT or 
\mjj bin are treated as uncorrelated within a channel and correlated across channels. 
The uncertainties from the closure tests are treated as uncorrelated within and across the different final states.

\begin{figure}[tbh]
   \centering
    \includegraphics[width=.45\textwidth]{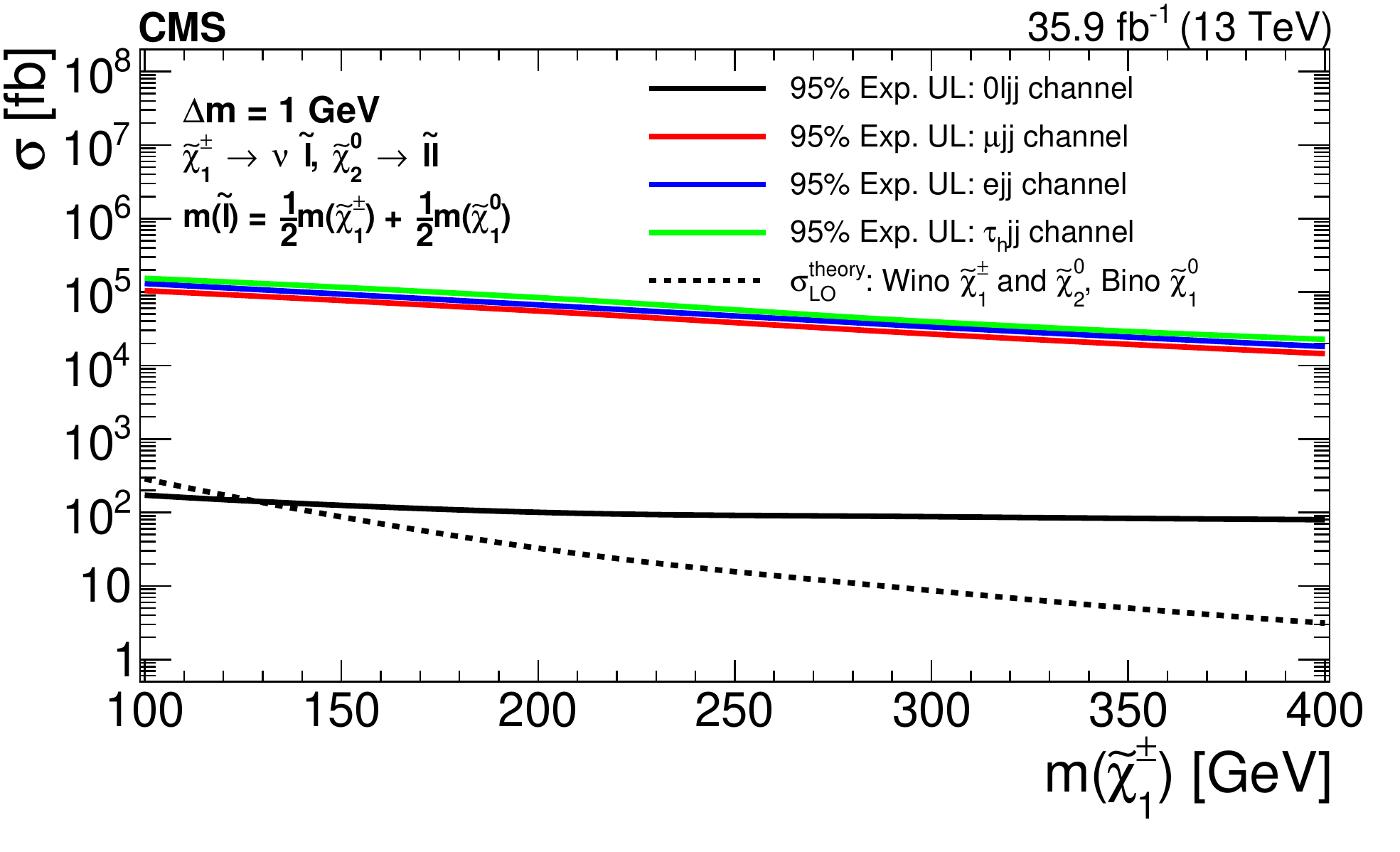}
    \includegraphics[width=.45\textwidth]{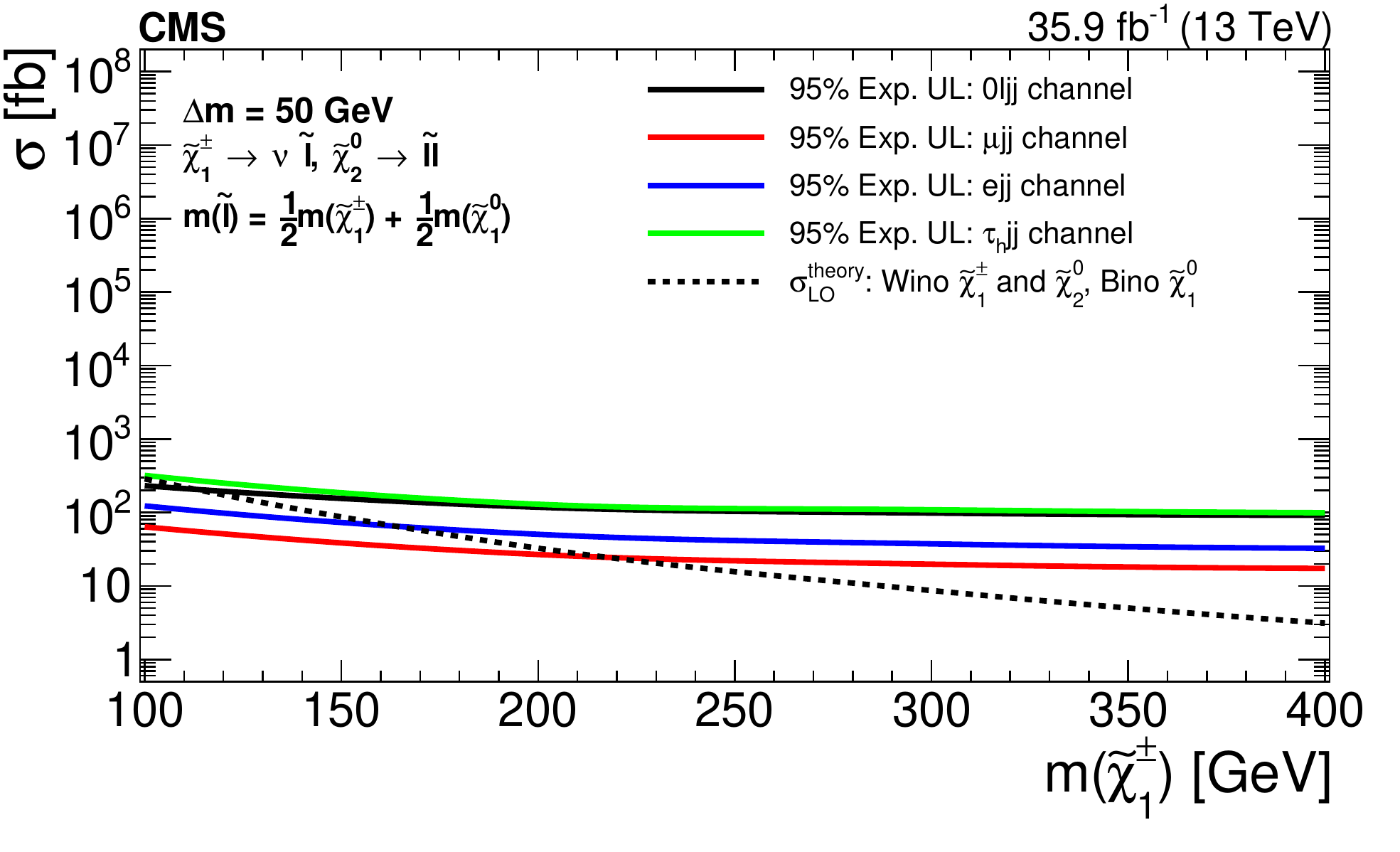}\\
    \includegraphics[width=.45\textwidth]{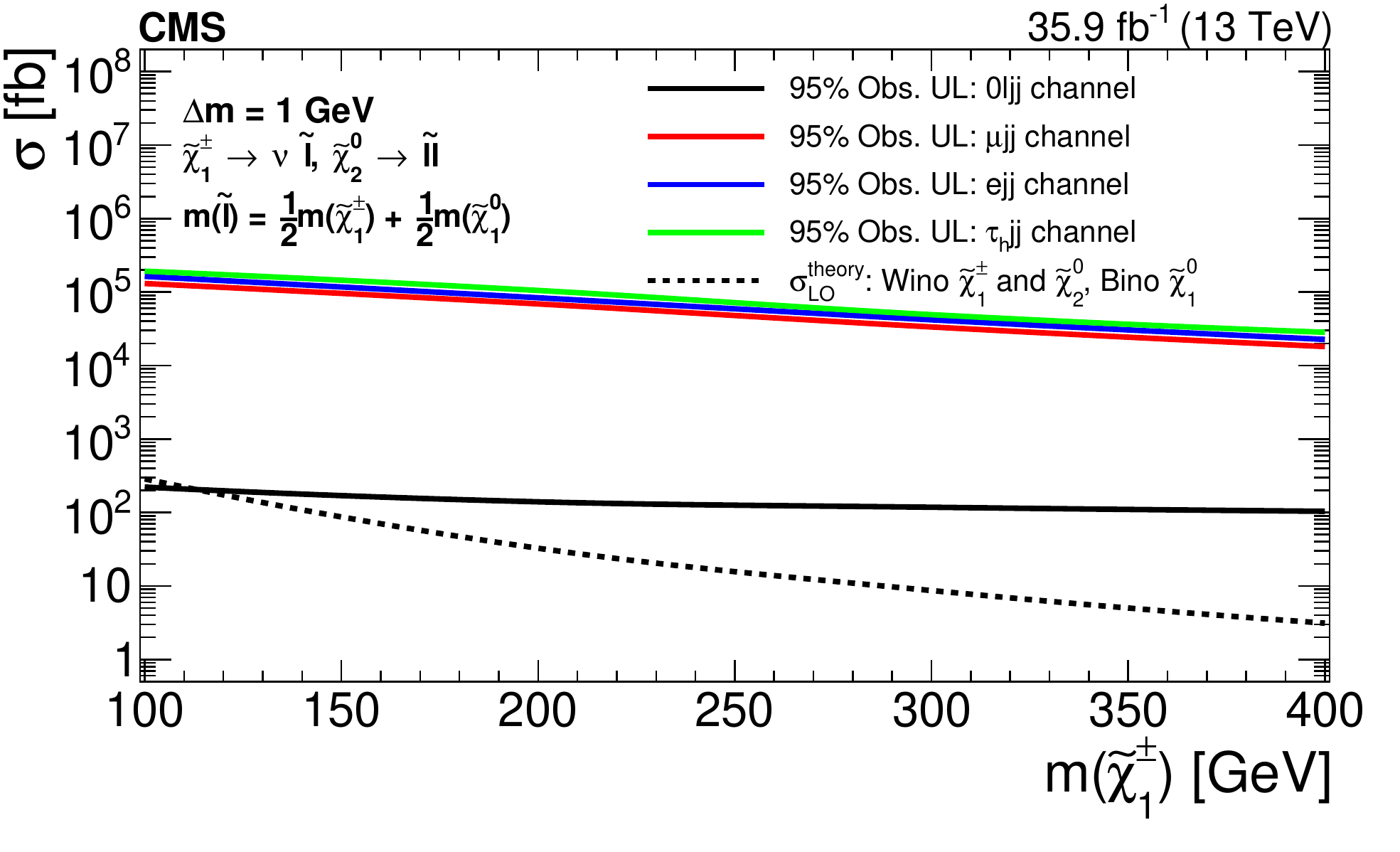}
    \includegraphics[width=.45\textwidth]{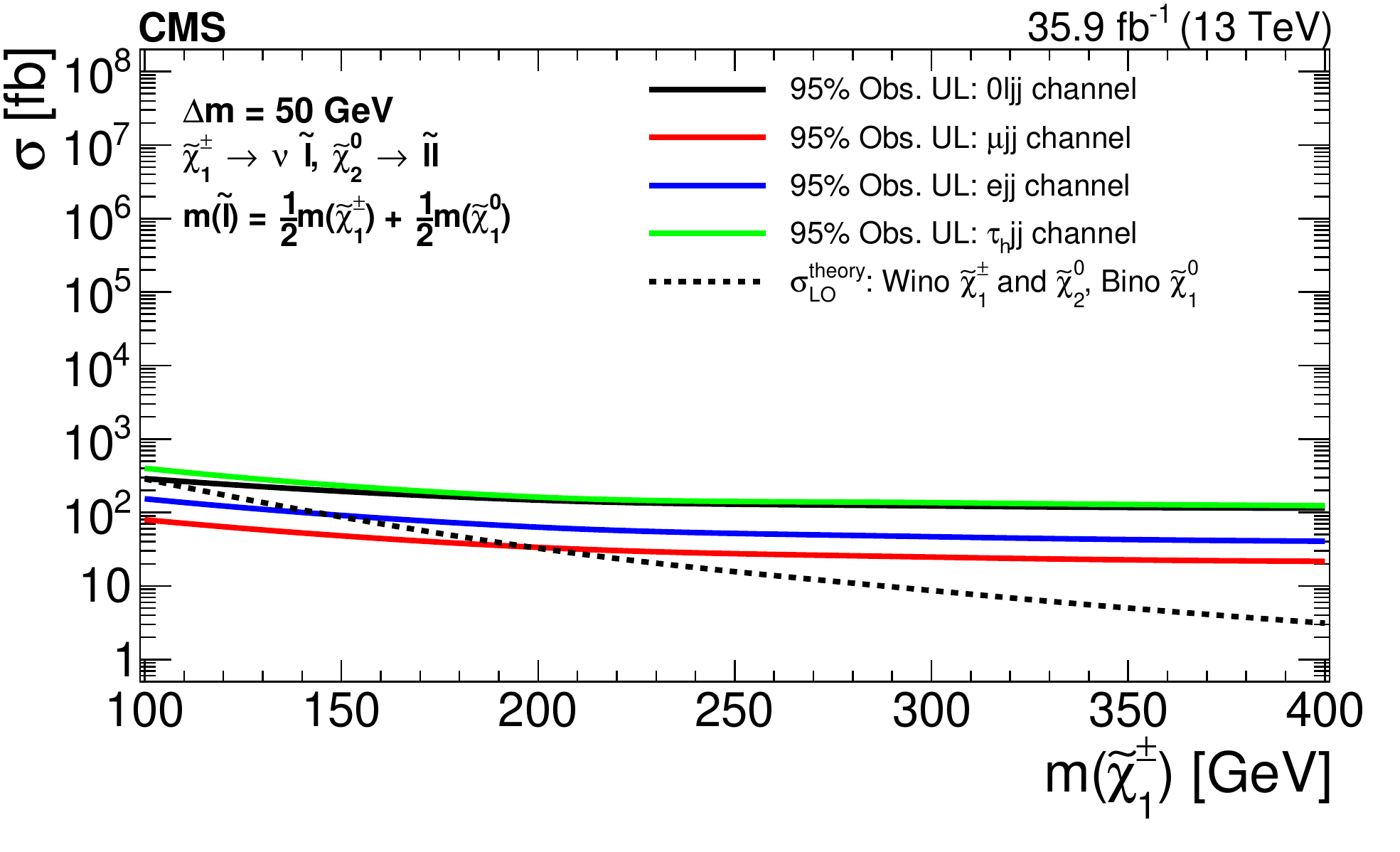}
     \caption{Combined 95\% \CL UL on the cross section as a function of $m_{\PSGczDt}=m_{\PSGcpmDo}$. The results correspond to $\Delta m=1\GeV$ (left) and $\Delta m=50\GeV$ (right) mass
gaps between the chargino and the lightest neutralino in the light slepton model. The top row shows the expected limits, and the bottom row shows the observed limits.}
 \label{fig:combined1DLimit}
\end{figure}

\begin{figure}
  \centering
  {\includegraphics[width=0.48\textwidth]{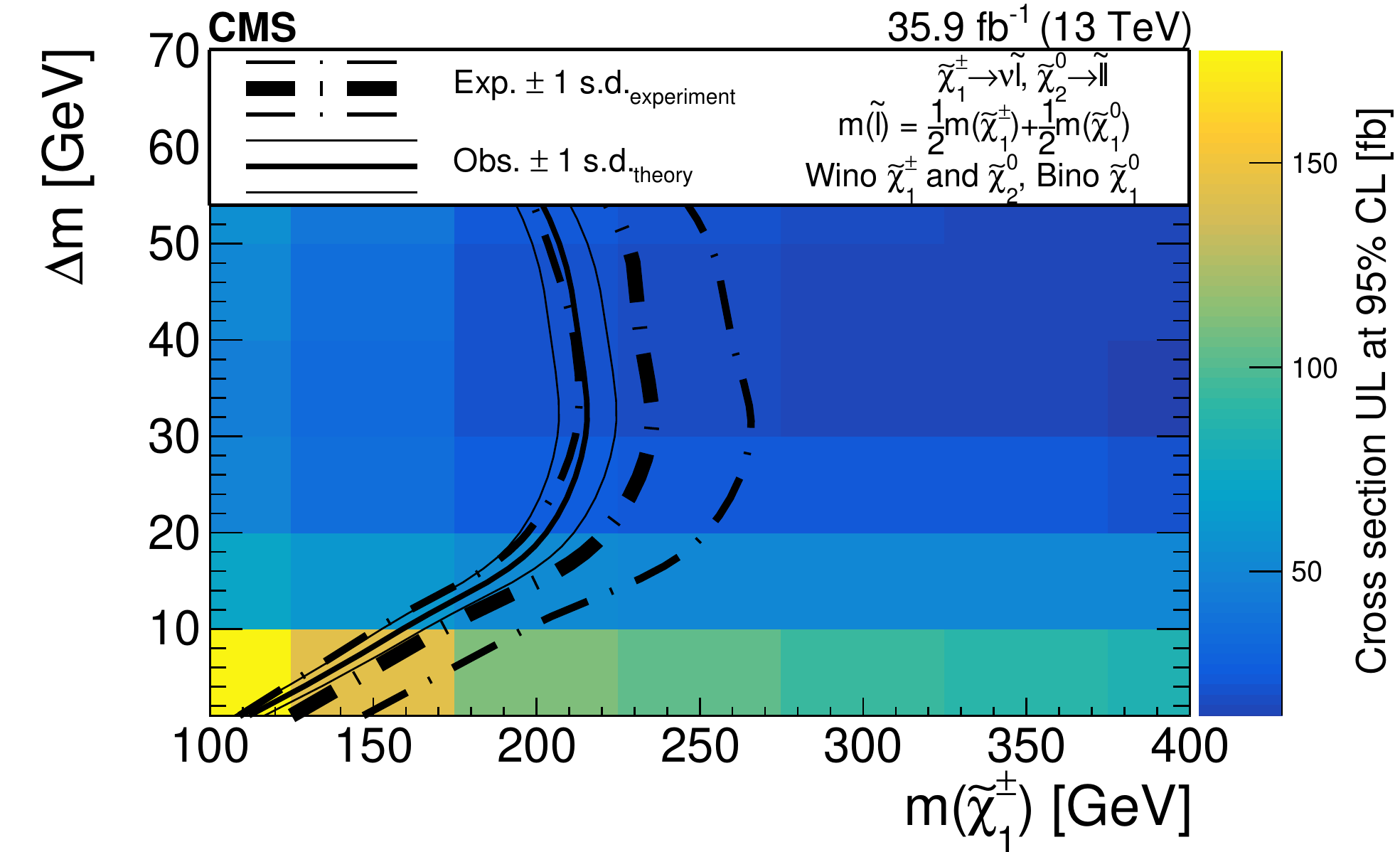}}
  {\includegraphics[width=0.48\textwidth]{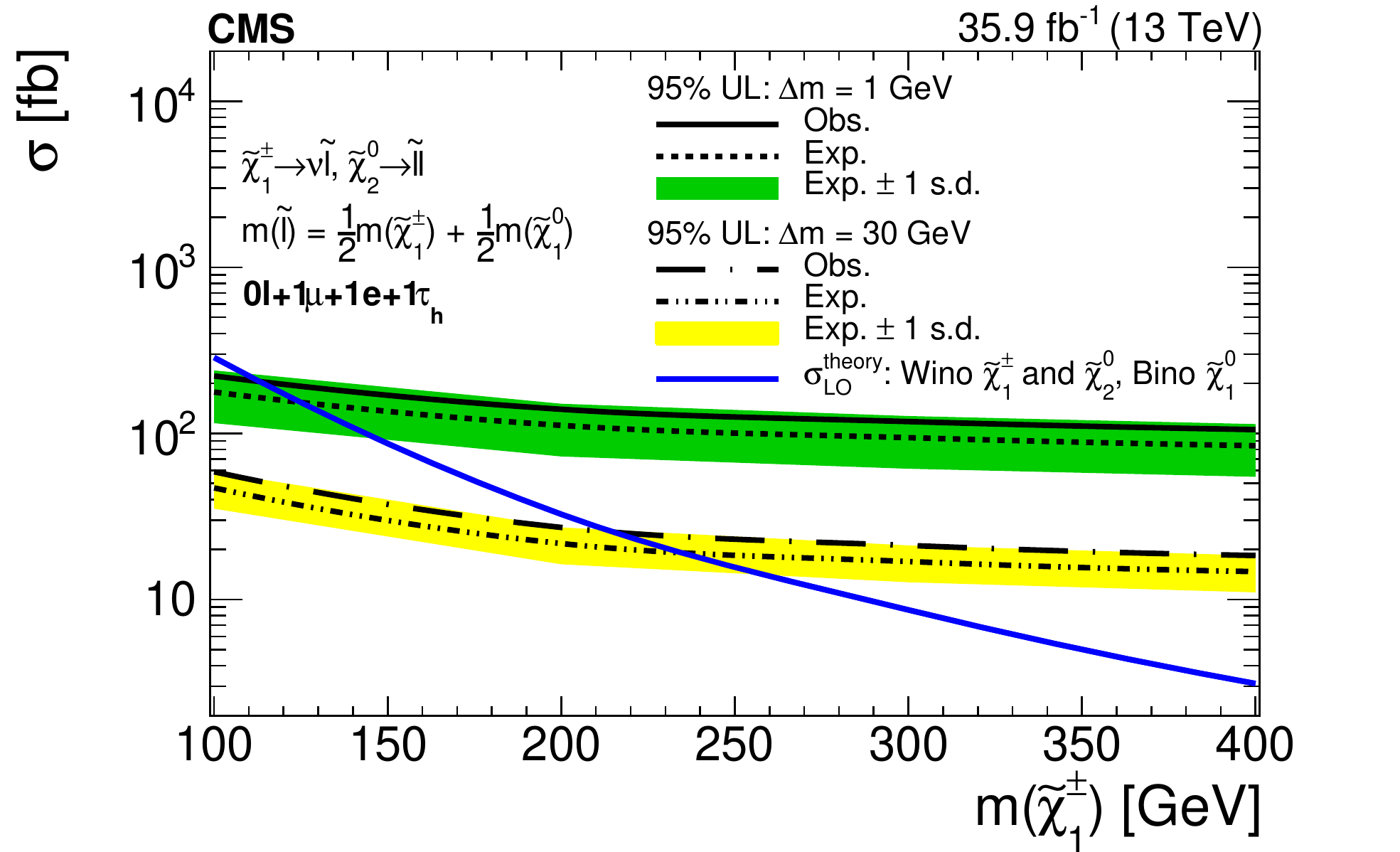}}

  \caption{(Left) Expected and observed 95\% confidence level upper limit (UL) on the signal cross section as a function of $m(\PSGcpmDo)$ and $\Delta m$,
assuming the light slepton model with slepton mass defined as the average of the \PSGczDt and \PSGcpmDo masses, $x_{\widetilde{\ell}}=0.5$. The lower left edge of each bin represents the 
\{$m(\PSGcpmDo)$,$\Delta m$\} combination used to calculate the UL on the signal cross section. For example, the lowest and leftmost bin corresponds to the UL on the signal cross section for the 
scenario with $m(\PSGcpmDo) = 100$\GeV and $\Delta m = 1$\GeV. (Right) Combined 95\% \CL UL 
on the cross section as a function of $m_{\PSGczDt}=m_{\PSGcpmDo}$, for $\Delta m=1\GeV$ and $\Delta m=30\GeV$ mass gaps between the chargino and the neutralino, assuming the light slepton model.}
    \label{fig:2Dlimit}
\end{figure}

Figure~\ref{fig:combined1DLimit} shows the expected and observed limits as well as the theoretical cross section as functions of $m_{\PSGcpmDo}$ for
the $\Delta m = 1$ and 50\GeV assumptions in the light slepton model.
For the smallest value of $\Delta m = 1\GeV$, the $0\ell \mathrm{jj}$ channel provides the best sensitivity, while the VBF
soft-$\Pe$ and soft-$\mu$ channels provide the best sensitivity for the larger mass gap scenario with $\Delta m = 50\GeV$.
The four channels are combined and the results are presented in Fig.~\ref{fig:2Dlimit}.
Figure~\ref{fig:2Dlimit} (left) shows the 95\% \CL UL on the signal cross section,
as a function of $m(\PSGcpmDo)$ and $\Delta m$, assuming $x_{\widetilde{\ell}}=0.5$.
Figure~\ref{fig:2Dlimit} (right) shows the 95\% \CL UL on the signal cross section,
as a function of $m(\PSGcpmDo)$, for two fixed $\Delta m$ values of 1 and 30\GeV, and assuming $x_{\widetilde{\ell}}=0.5$.
The signal acceptance and mass shape are evaluated for
each \{$m(\PSGcpmDo), \Delta m$\} combination and used in the limit calculation procedure described above.
For the $\Delta m = \{1,10,30,50\}\GeV$ assumption, the combination of the four channels results in an observed (expected) exclusion on the \PSGczDt and \PSGcpmDo gaugino masses below
\{$112,159,215,207$\} (\{$125,171,235,228$\})\GeV.
For the compressed mass spectrum scenarios with $1 \le \Delta m \le 30\GeV$, the bounds on the \PSGczDt and \PSGcpmDo gaugino masses are the most stringent to date.

It is noted that for the $1 < \Delta m < 10\GeV$ mass gaps considered in this analysis, the exclusions on $m(\PSGcpmDo)$ do not depend on the
assumption that a light slepton exists (\ie $m(\PSGc^{\pm}_{0}) < m_{\widetilde{\ell}} < m(\PSGcpmDo)$). For $1 < \Delta m < 10\GeV$,
the signal acceptance for the $\PW\PZ$ model is similar to the signal acceptance for the light slepton model.
For example, Fig.~\ref{fig:SRPlots} (lower right) shows the expected \mjj
signal distribution when the decays of the charginos and neutralinos proceed via {\PW} and {\PZ} bosons, resulting in a similar shape and normalization as the expectation for the
light slepton scenario. However, for increasing $\Delta m$ values where the $1\ell \mathrm{jj}$ channels dominate the sensitivity, the exclusions on $m(\PSGcpmDo)$ in the $\PW\PZ$ model
are less stringent than the ones in the light slepton model. This difference is a result of the lower branching ratio of \PSGcpmDo and
\PSGczDt to leptonic final states in the $\PW\PZ$ model.

\begin{figure}
  \centering
  {\includegraphics[width=0.48\textwidth]{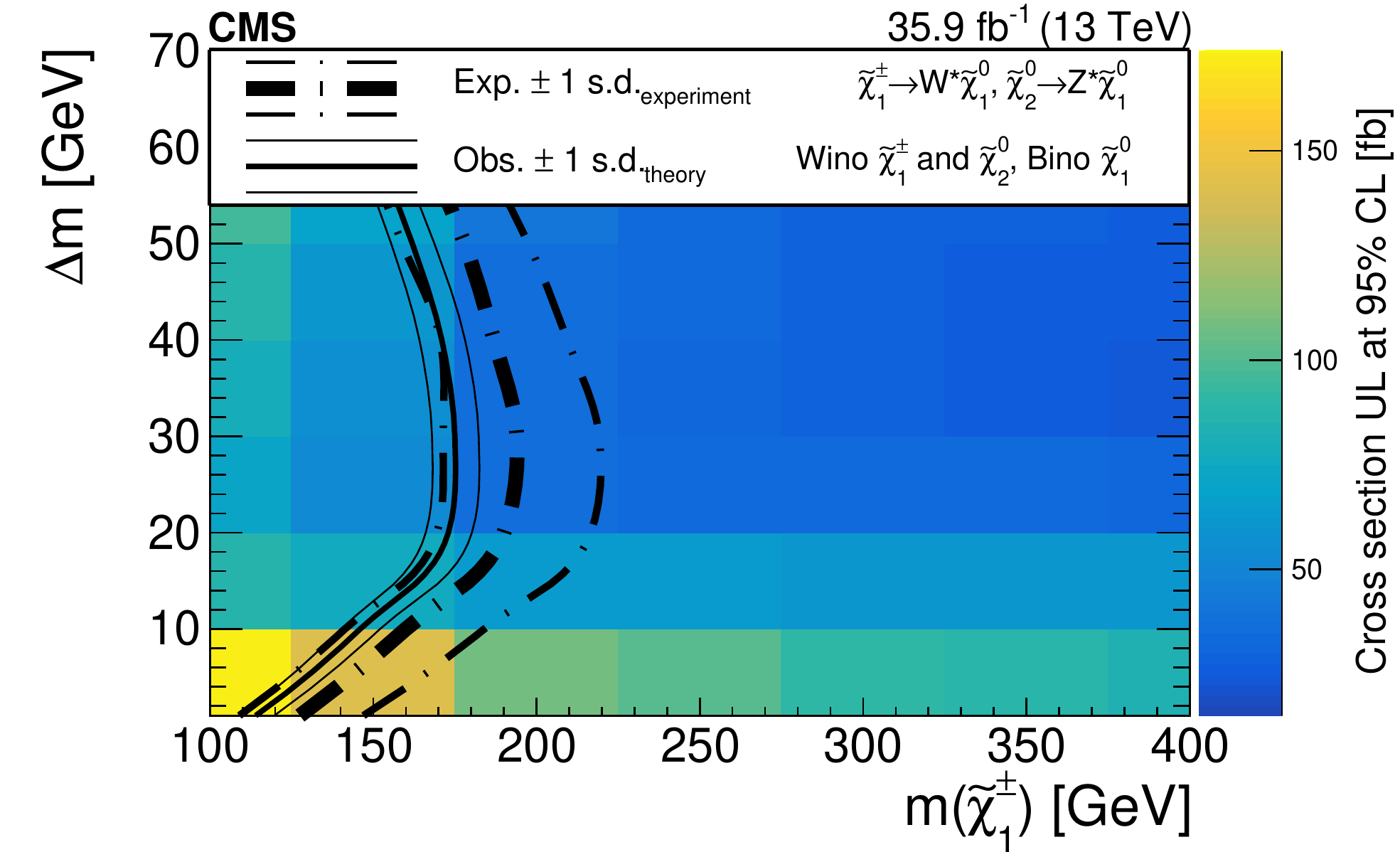}}
  {\includegraphics[width=0.48\textwidth]{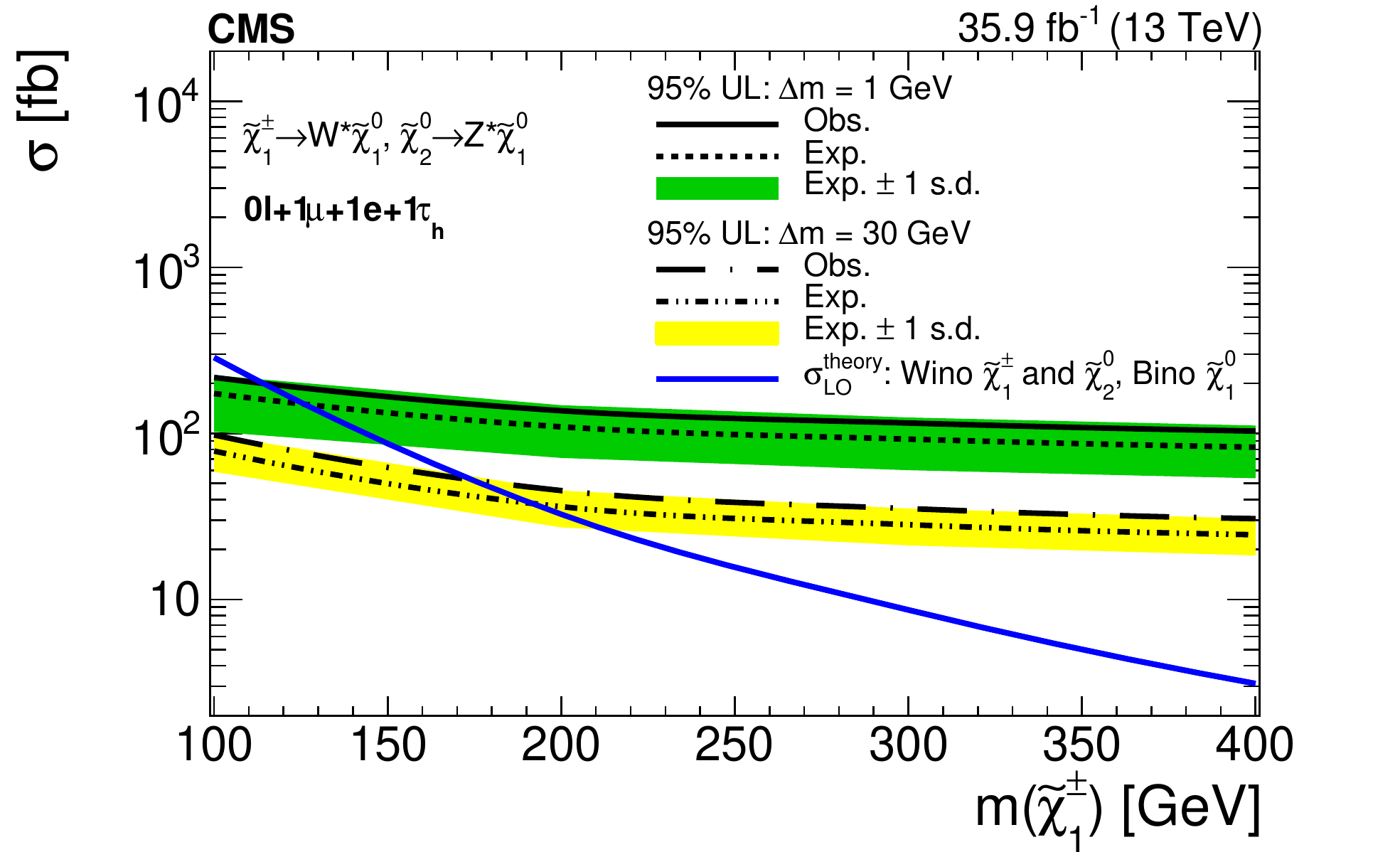}}
  \caption{(Left) Expected and observed 95\% confidence level upper limit (UL) on the signal cross section as a function of $m(\PSGcpmDo)$ and $\Delta m$,
assuming the \PSGcpmDo and \PSGczDt decays proceed via $\PW^{*}$ and $\PZ^{*}$. The lower left edge of each bin represents the 
\{$m(\PSGcpmDo)$,$\Delta m$\} combination used to calculate the UL on the signal cross section. For example, the lowest and leftmost bin corresponds to the UL on the signal cross section for the
scenario with $m(\PSGcpmDo) = 100$\GeV and $\Delta m = 1$\GeV. (Right) The 95\% \CL UL on the cross section as a function of $m_{\PSGczDt}=m_{\PSGcpmDo}$, for $\Delta 
m=1\GeV$ and $\Delta m=30\GeV$ mass gaps between the chargino and the neutralino, after combining 0 lepton and 1 lepton channels, assuming the \PSGcpmDo and \PSGczDt
decays proceed via $\PW^{*}$ and $\PZ^{*}$.}
    \label{fig:2DlimitWZ}
\end{figure}

Figure~\ref{fig:2DlimitWZ} (left) shows the 95\% \CL UL on the signal cross section,
as a function of $m(\PSGcpmDo)$ and $\Delta m$, assuming the $\PW\PZ$ model.
Figure~\ref{fig:2DlimitWZ} (right) shows the 95\% \CL UL on the signal cross section,
as a function of $m(\PSGcpmDo)$, for two fixed $\Delta m$ values of 1 and 30\GeV, and assuming the $\PW\PZ$ model.
For the $\Delta m =$ \{$1,10,30,50$\}\GeV assumption, the combination of the four channels results in an observed (expected) exclusion on the \PSGczDt and \PSGcpmDo gaugino masses below
\{$112,146,175,162$\} (\{$125,160,194,178$\})\GeV.
For the compressed mass spectrum scenarios with $1 \le \Delta m < 3\GeV$ and $25 \le \Delta m <50\GeV$,
the bounds on the \PSGczDt and \PSGcpmDo gaugino masses in the
$\PW\PZ$ model are also the most stringent to date, surpassing the bounds from the LEP experiments~\cite{Heister:2003zk,Heister:2002mn,Abbiendi:2002vz,Acciarri:2000wy}.

\section{Summary}
\label{sec:sum}

A search is presented for noncolored supersymmetric particles produced in the vector boson fusion (VBF) topology
using data corresponding to an integrated luminosity of 35.9\fbinv collected in 2016 with the CMS detector
in proton-proton collisions at $\sqrt{s}=13\TeV$.
The search utilizes events in four different channels depending on the number and type of leptons: $0\ell \mathrm{jj}$, $\Pe \mathrm{jj}$, $\mu \mathrm{jj}$, and $\tauh \mathrm{jj}$, where $\tauh$ denotes a
hadronically decaying {\Pgt} lepton.
While Ref.~\cite{SUS14019} reported a search using the VBF dijet topology with a zero-lepton final state in proton-proton collision data at $\sqrt{s}=8\TeV$, this is the first
search for the compressed electroweak supersymmetry (SUSY) sector using the $0\ell \mathrm{jj}$ final state.
This is also the first search for SUSY in the VBF topology with single soft-lepton final states.
The VBF topology requires two well-separated jets that appear in opposite hemispheres, with large invariant mass \mjj.
The observed \mjj and transverse mass $\mT(\ell,\ptmiss)$ distributions do not reveal any evidence for new physics.
The results are used to exclude a range of \PSGcpmDo and \PSGczDt gaugino masses.
For a compressed mass spectrum scenario, in which $\Delta m \equiv m(\PSGcpmDo) -m(\PSGczDo) = 1$ (30)\GeV and in which \PSGcpmDo and \PSGczDt branching fractions to light sleptons are 100\%,
\PSGcpmDo and \PSGczDt masses up to 112 (215)\GeV are excluded at 95\% \CL. For the scenario where the sleptons are too heavy and decays of the charginos and neutralinos proceed via $\PW^{*}$
and $\PZ^{*}$ bosons, \PSGcpmDo and \PSGczDt masses up to 112 (175)\GeV are excluded at 95\% \CL for $\Delta m = 1$ (30)\GeV.
While many previous studies at the LHC have focused on strongly coupled supersymmetric particles, including searches for charginos and neutralinos produced in gluino or squark decay chains, and a
number of studies have presented limits on the Drell--Yan production of charginos and neutralinos, this analysis obtains the most stringent limits to date on the production of charginos and
neutralinos decaying to leptons in compressed mass spectrum scenarios defined by
the mass separation $1 \le \Delta m < 3\GeV$ and $25 \le \Delta m < 50\GeV$.

\begin{acknowledgments}
We congratulate our colleagues in the CERN accelerator departments for the excellent performance of the LHC and thank the technical and administrative staffs at CERN and at other CMS institutes for their contributions to the success of the CMS effort. In addition, we gratefully acknowledge the computing centers and personnel of the Worldwide LHC Computing Grid for delivering so effectively the computing infrastructure essential to our analyses. Finally, we acknowledge the enduring support for the construction and operation of the LHC and the CMS detector provided by the following funding agencies: BMBWF and FWF (Austria); FNRS and FWO (Belgium); CNPq, CAPES, FAPERJ, FAPERGS, and FAPESP (Brazil); MES (Bulgaria); CERN; CAS, MoST, and NSFC (China); COLCIENCIAS (Colombia); MSES and CSF (Croatia); RPF (Cyprus); SENESCYT (Ecuador); MoER, ERC IUT, PUT and ERDF (Estonia); Academy of Finland, MEC, and HIP (Finland); CEA and CNRS/IN2P3 (France); BMBF, DFG, and HGF (Germany); GSRT (Greece); NKFIA (Hungary); DAE and DST (India); IPM (Iran); SFI (Ireland); INFN (Italy); MSIP and NRF (Republic of Korea); MES (Latvia); LAS (Lithuania); MOE and UM (Malaysia); BUAP, CINVESTAV, CONACYT, LNS, SEP, and UASLP-FAI (Mexico); MOS (Montenegro); MBIE (New Zealand); PAEC (Pakistan); MSHE and NSC (Poland); FCT (Portugal); JINR (Dubna); MON, RosAtom, RAS, RFBR, and NRC KI (Russia); MESTD (Serbia); SEIDI, CPAN, PCTI, and FEDER (Spain); MOSTR (Sri Lanka); Swiss Funding Agencies (Switzerland); MST (Taipei); ThEPCenter, IPST, STAR, and NSTDA (Thailand); TUBITAK and TAEK (Turkey); NASU and SFFR (Ukraine); STFC (United Kingdom); DOE and NSF (USA).

\hyphenation{Rachada-pisek} Individuals have received support from the Marie-Curie program and the European Research Council and Horizon 2020 Grant, contract Nos.\ 675440, 752730, and 765710 (European Union); the Leventis Foundation; the A.P.\ Sloan Foundation; the Alexander von Humboldt Foundation; the Belgian Federal Science Policy Office; the Fonds pour la Formation \`a la Recherche dans l'Industrie et dans l'Agriculture (FRIA-Belgium); the Agentschap voor Innovatie door Wetenschap en Technologie (IWT-Belgium); the F.R.S.-FNRS and FWO (Belgium) under the ``Excellence of Science -- EOS" -- be.h project n.\ 30820817; the Beijing Municipal Science \& Technology Commission, No. Z181100004218003; the Ministry of Education, Youth and Sports (MEYS) of the Czech Republic; the Lend\"ulet (``Momentum") Program and the J\'anos Bolyai Research Scholarship of the Hungarian Academy of Sciences, the New National Excellence Program \'UNKP, the NKFIA research grants 123842, 123959, 124845, 124850, 125105, 128713, 128786, and 129058 (Hungary); the Council of Science and Industrial Research, India; the HOMING PLUS program of the Foundation for Polish Science, cofinanced from European Union, Regional Development Fund, the Mobility Plus program of the Ministry of Science and Higher Education, the National Science Center (Poland), contracts Harmonia 2014/14/M/ST2/00428, Opus 2014/13/B/ST2/02543, 2014/15/B/ST2/03998, and 2015/19/B/ST2/02861, Sonata-bis 2012/07/E/ST2/01406; the National Priorities Research Program by Qatar National Research Fund; the Ministry of Science and Education, grant no. 3.2989.2017 (Russia); the Programa Estatal de Fomento de la Investigaci{\'o}n Cient{\'i}fica y T{\'e}cnica de Excelencia Mar\'{\i}a de Maeztu, grant MDM-2015-0509 and the Programa Severo Ochoa del Principado de Asturias; the Thalis and Aristeia programs cofinanced by EU-ESF and the Greek NSRF; the Rachadapisek Sompot Fund for Postdoctoral Fellowship, Chulalongkorn University and the Chulalongkorn Academic into Its 2nd Century Project Advancement Project (Thailand); the Welch Foundation, contract C-1845; and the Weston Havens Foundation (USA).
\end{acknowledgments}

\bibliography{auto_generated}
\cleardoublepage \appendix\section{The CMS Collaboration \label{app:collab}}\begin{sloppypar}\hyphenpenalty=5000\widowpenalty=500\clubpenalty=5000\vskip\cmsinstskip
\textbf{Yerevan Physics Institute, Yerevan, Armenia}\\*[0pt]
A.M.~Sirunyan, A.~Tumasyan
\vskip\cmsinstskip
\textbf{Institut für Hochenergiephysik, Wien, Austria}\\*[0pt]
W.~Adam, F.~Ambrogi, E.~Asilar, T.~Bergauer, J.~Brandstetter, M.~Dragicevic, J.~Erö, A.~Escalante~Del~Valle, M.~Flechl, R.~Frühwirth\cmsAuthorMark{1}, V.M.~Ghete, J.~Hrubec, M.~Jeitler\cmsAuthorMark{1}, N.~Krammer, I.~Krätschmer, D.~Liko, T.~Madlener, I.~Mikulec, N.~Rad, H.~Rohringer, J.~Schieck\cmsAuthorMark{1}, R.~Schöfbeck, M.~Spanring, D.~Spitzbart, W.~Waltenberger, J.~Wittmann, C.-E.~Wulz\cmsAuthorMark{1}, M.~Zarucki
% ...
\vskip\cmsinstskip
% ...
1:  Also at Vienna University of Technology, Vienna, Austria\\
% ...
\end{sloppypar}
\end{document}